\documentstyle[bbm,amsfonts, epsfig,12pt]{article}

\newcommand {\eqref} [1] {(\ref {#1})}
\newcommand {\slsh} [1] {\not{\hbox{\kern-2pt${#1}$}}}

\newcommand {\beq} {\begin{equation}}    
\newcommand {\eeq} {\end{equation}}
  \newcommand {\ber}{\begin{eqnarray*}}
  \newcommand {\eer} {\end{eqnarray*}}
\newcommand {\beqn}{\begin{eqnarray}}
  \newcommand {\eeqn} {\end{eqnarray}}

\newcommand{\Dslash}{\,{\raise.15ex\hbox{/}\mkern-12mu D}}
\newcommand{\Tr}{{\rm Tr}\,}
\newcommand{\gsim}{\lower.7ex\hbox{$
\;\stackrel{\textstyle>}{\sim}\;$}}
\newcommand{\lsim}{\lower.7ex\hbox{$
\;\stackrel{\textstyle<}{\sim}\;$}}

\def\href#1#2{#2}	

\def\coeff#1#2{{\textstyle {\frac {#1}{#2}}}}
\def\half{\coeff 12}

\def\N{{\mathcal N}}

\def\S_1{{\widetilde {S_1}}}

\def\R{{\mathbb R}}

\def\tr{{\rm Tr}}

\def\Z{{\mathbb Z}}

\def\Dslash{{\rlap{\raise 1pt \hbox{$\>/$}}D}}

\begin{document}
\begin{titlepage}
\begin{flushright}{UMN-TH-2704/08\,,  \,\,\, FTPI-MINN-08/25\,,  \,\, SLAC-PUB-13372
\\
8-15/08 
}
\end{flushright}
\vskip 0.2cm

\centerline{{\Large \bf  On  Yang--Mills  Theories with Chiral Matter  }}

\vspace{2mm}

\centerline{{\Large \bf   at Strong Coupling  }}

\vskip 1cm
\centerline{\large  M. Shifman${}^{a,b}$  and Mithat \"{U}nsal ${}^{c,d}$}

\vskip 0.2cm

\centerline{${}^a$   \footnotesize\it William I. Fine Theoretical Physics Institute,}
\centerline{\footnotesize\it University of Minnesota, Minneapolis, MN 55455, USA}

\vskip 0.2cm

\begin{center}
${}^b$ {\footnotesize\it 
Laboratoire de Physique Th\'eorique\footnote{Unit\'e
Mixte de Recherche du CNRS,  (UMR 8627).}
Universit\'e de Paris-Sud XI\\
B\^atiment 210, 
F-91405 Orsay C\'edex, FRANCE
}
\end{center}
\vskip 0.2cm

\centerline{${}^c$ \footnotesize\it SLAC, Stanford 
University, Menlo Park, CA 94025, USA}
\vskip 0.2cm
\centerline{${}^d$ \footnotesize\it  Physics Department, 
Stanford University, Stanford, CA 94305, USA }

\begin{abstract}
Strong coupling dynamics of Yang--Mills theories with chiral fermi\-on content remained 
largely elusive despite much effort over the years. In this work, we propose 
a dynamical framework in which we can address non-perturbative properties of  
chiral, non-supersymmetric  gauge theories, in particular, chiral quiver theories  
on $S_1 \times R_3 $.  Double-trace deformations are used to 
stabilize the center-symmetric vacuum. This allows one to
smoothly connect small-$r(S_1)$ to large-$r(S_1)$ physics ($R_4$ 
is the limiting case)
where the double-trace deformations are switched off. In particular,  
occurrence of the 
mass gap in the gauge sector and linear confinement  due to bions 
are analytically
demonstrated. We find the pattern of the chiral symmetry realization
which depends on the structure of the ring operators, a novel
class of topological excitations.  

The deformed chiral theory, unlike the undeformed one, satisfies  
volume independence down to arbitrarily small volumes 
(a working  Eguchi--Kawai reduction) in the large $N$ limit.  
This equivalence, may open new perspectives on strong 
coupling chiral gauge theories on $R_4$.  

\end{abstract}

\end{titlepage}

\tableofcontents

\newpage

\section{Introduction}
\label{intro}

Recent advances in strongly coupled Yang--Mills theories,
both analytic and numerical, are quite spectacular.
At the same time, our knowledge of (ano\-maly free) Yang--Mills theories
with chiral matter remains at a rudimentary level.
All existing methods of exploration  
fail in the chiral case.  Lattice techniques at the moment do not lead to a manifestly 
gauge invariant formulation of the non-Abelian chiral gauge theories. (For progress in this 
direction, see  \cite{Neuberger:2001nb, Luscher:2000hn,Golterman:2004qv,Poppitz:2007tu} and references therein.) 
Even if a satisfactory gauge invariant lattice formulation was constructed, the  
well known problem of  complex fermion determinants would render numerical simulations impractical. 
Analytic arsenals of theorists dealing with chiral matter at strong coupling
are poor, to put it mildly. The 't Hooft matching \cite{hooft} is a useful tool, generally speaking.
However, since we will mostly focus on 
$Z_K$-orbifold theories with no continuous global axial symmetries, 
the 't Hooft matching \cite{hooft} is not applicable in this case. 
AdS/QCD modeling  and string theory techniques (so far) do not provide insights  
into the strong coupling chiral dynamics either. 

In essence, the only fact which can be considered established is
the {\em perturbative} equivalence of the $Z_K$ orbifold theories  
at large $N$ to  supersymmetric SU$(KN)$ Yang--Mills theory, 
with an appropriate rescaling of the gauge couplings
\cite{Kachru:1998ys}. This  equivalence 
does {\em not} extend beyond perturbation 
theory in the chiral case $(K \geq 3)$ due to spontaneous breaking of the chiral symmetry 
in the parent theory  \cite{Kovtun:2005kh} (which is used in 
the projection).\footnote{Discussion of the non-perturbative fate of planar equivalence for
$Z_K$ orbifolds ($K\geq 3$) was initiated in Refs.~\cite{Schmaltz:1998bg,stras1}.} 
Thus, the above planar equivalence tells us nothing about such basic features
of the theory as the vacuum degeneracy/nondegeneracy, patterns of the
discrete chiral symmetry breaking and so on, let alone
the spectrum of composite colorless hadrons. 

In this paper we discuss dynamics of non-Abelian
gauge theories with chiral fermion sectors at strong coupling
applying and developing ideas suggested in \cite{Shifman:2008ja}. 
Our primary target is the so-called $Z_K$ orbifold theories at $K\geq 3$. 
These theories are obtained from  supersymmetric SU$(KN)$ Yang--Mills theory
by  $Z_K$ orbifold projection. The gauge group is [SU$(N)]^K$, i.e 
they contain $K$ gluon sectors which are connected to each other
only through bifundamental {\em Weyl} fermions of the type 
$ 
 \psi_J \sim (1, \ldots, N_J , \overline{N}_{J+1}, \ldots 1) 
$ 
which transform in the  fundamental representation  of gauge factor  
 SU$(N)_J$  and anti-fundamental of  SU$(N)_{J+1}$. Here
$J=1,2, ...,K$ labels the gauge factors.  These theories are also known as ``quiver."

If $K\geq 3$ the mass term for fermions cannot be introduced since there are no gauge invariant bifermion operators. Thus, these theories are genuinely chiral. They have no internal anomalies
and are well-defined. It is clear that understanding of strong coupling gauge dynamics
is impossible without answering the question of their dynamical behavior, of which next-to-nothing is known.  Understanding such theories also carries a  phenomenological interest. 
Strongly coupled chiral gauge theories may be relevant for TeV-scale physics,
in particular, bearing responsibility for the electro-weak symmetry breaking, and 
fermion masses.\footnote{Early discussions of possible patterns of
chiral symmetry breaking in Yang--Mills theories with fermion fields
in various representations and consequences for particle spectra
can be found in Refs.~\cite{Dimopoulos:1980hn,Peskin:1982mu}.}

Our goal  is to understand non-perturbative 
dynamics of chiral gauge theories in the continuum limit
in a locally four-dimensional setting.  
Currently,  there  exists no controllable
{\em dynamical framework} which one could use 
to address non-perturbative aspects of chiral  theories. We suggest one.
As a matter of fact, the method designed and applied in vector-like
gauge theories with one flavor 
Ref.~\cite{Shifman:2008ja} (see also \cite{Armoni:2007kd}) and 
in pure Yang--Mills theory \cite{Unsal:2008ch} can be adjusted to become an analytical
tool in chiral gauge theories. 

The above-mentioned method has several key elements. First, instead of considering
Yang--Mills theories on $R_4$ we compactify one of the dimensions replacing $R_4$ by
$R_3\times S_1$. Then we analyze the theory on $R_3\times S_1$. At small
$r(S_1)$ the theory can be made weakly coupled, with full control over non-perturbative
effects. We perform the so-called
``double-trace"
deformation of the theory in the small-$r(S_1)$ domain.
It is designed in a such a way that the small-$r(S_1)$ theory becomes 
continuously connected to the undeformed theory on  $R_4$.  
 If the deformation was
not performed, we would encounter 
phase transitions on the way from  small to large $r(S_1)$.
Physics of these two domains would be different, 
and studying the theory at small $r(S_1)$ would not tell us much about the large-$r(S_1)$ dynamics.
With an appropriately chosen deformation
we can avoid unwanted phase transitions 
ensuring qualitative validity of small-$r(S_1)$ results at large $r(S_1)$. 
The deformed theories are labeled by asterisk, such as YM*.
   
In fact, the role of the double-trace deformation
is to stabilize the center symmetry in the   small $r(S_1)$ 
regime.\footnote{A double-trace deformation which is insufficient to 
completely stabilize the center symmetry will result,
generally speaking, in novel phases with a partially 
broken center symmetry~\cite{Ogilvie:2007tj, Myers:2007vc}.  
The constructions in  Refs.~\cite{Shifman:2008ja,Unsal:2008ch}   
avoid the presence of such exotic phases by crafting a sufficient deformation.  
One can infer from existing lattice simulations (see e.g.~Fig.~1 in
Ref.~\cite{Ogilvie:2007tj}) that a
sufficiently large deformation stabilizes 
the center symmetry  at any value of the bare lattice coupling.
This numerically confirms our proposal. An earlier example of 
QCD-like gauge theories with unbroken center symmetry  
can be found in~\cite{Kovtun:2007py}.} 
Although
this deformation is essential at small $r(S_1)$, it
can be switched off at large $r(S_1)$.  The fact that
at  small $r(S_1)$ the center symmetry is not spontaneously broken guarantees continuity,
 at least in the sense of Polyakov's order parameter. Moreover, 
the theory is at weak coupling at small $r(S_1)$. Everything is analytically calculable.
The theory has a rich non-perturbative sector which can be treated quasiclassically.
It is populated by instanton-monopoles of two types
('t Hooft--Polyakov and Kaluza--Klein) and a variety of composite topological objects
built of the above instanton-monopoles. Such  composites will be referred to as 
 instanton-monopole molecules.  Mathematically, these correspond to  magnetic or topological  flux carrying operators. In the  
 {\it weak} coupling regime, a center-symmetric configuration of the  (untraced) Polyakov line  
 behaves as an adjoint Higgs field with a non-vanishing expectation value. 
 The non-Abelian gauge symmetry is spontaneously
 broken; the gauge structure reduces to the maximal Abelian 
 subgroup,
\beq
{\rm SU}(N) \to \left[ {\rm U}(1)^{N-1}\right]\,.
\label{pat1}
\eeq
The off-diagonal ``$W$ bosons" become heavy at small $r(S_1)$, and 
play no role in the infrared (IR) dynamics.
The diagonal photons remain massless in perturbation theory.

However, non-perturbatively all (dual) photons acquire mass terms
through the magnetic monopole-instantons \cite{Unsal:2008ch} 
or magnetic  bions \cite{Shifman:2008ja}, 
via the  Polyakov mechanism \cite{Polyakov:1976fu}. This results in formation 
of the flux ``tubes" (strings) in two spatial dimensions, guaranteeing
linear confinement. In addition, in QCD*-like theories, the   instanton-monopoles generate
bifermion vertices  leading to spontaneous breaking of the discrete chiral symmetry
and a vacuum degeneracy. 
The very same features --- linear confinement and spontaneous breaking of the discrete chiral symmetry --- are expected in these theories in the decompactification limit
$r(S_1)\to \infty$. Because of this, we argued \cite{Shifman:2008ja}
 the transition from small to large
$r(S_1)$ to be smooth in  one-flavor 
theories.\,\footnote{In extrapolating from small to large
$r(S_1)$ we, in fact, pass from Abelian to non-Abelian confinement.}
 If so, analytical results reliably obtained at
small $r(S_1)$ can be qualitatively extrapolated into the large-$r(S_1)$
domain.

The same strategy will be applied to chiral fermions,
in particular, in $Z_K$ orbifolds ($K\geq 3$), in which
at small $r(S_1)$ (i.e. at weak coupling)  the gauge symmetry reduces to 
\beq
[{\rm SU}(N)]^K \to \left[ {\rm U}(1)^{N-1}\right]^K\,.
\label{pat2}
\eeq
We calculate 
non-perturbative effects controlled by instanton-monopole molecules (flux operators) and then use these results to describe general features  in the decompactification limit, i.e. 
in chiral Yang--Mills theories at strong coupling. 
 For this construction to be valid it is important
that the $Z_K$ orbifold theories have no continuous axial global symmetries. 
 
Compared to QCD-like theories, in chiral theories we find
surprises.  As was mentioned, the monopole-instantons play a major role in small-$r(S_1)$ 
Abelian confinement regime \cite{Shifman:2008ja}.  The most surprising finding 
in the chiral quiver theories is that, despite the gauge symmetry breaking (\ref{pat2})  
in the small-$r(S_1)$ regime, the 
effect of monopole-instantons  identically vanishes!  If we 
denote the monopole-instanton action by $S_0$, 
\beq
S_0 =\frac{8\pi^2}{N\,g^2_{3+1}}\,,
\label{esnol}
\eeq
the leading contribution to the non-perturbative dynamics occurs at 
order $e^{-2S_0}$ via the magnetic bions. The leading terms   $e^{-S_0}$  
in the non-perturbative expansion cancels 
due to averaging over certain global symmetries.  There is also a plethora of non-perturbative 
flux operators appearing in the $e^{-S_0}$ expansion, which are neither monopoles 
nor instantons and some of which are special for chiral theories.  
 
Although our prime focus is
the $Z_K$   orbifolds with $K\geq 3$, we will briefly consider another class of
chiral gauge theories, namely, a 
single SU$(N)$ gauge group, with a chiral content, such as one Weyl fermion in the two-index antisymmetric (symmetric) representation supplemented by $N-4$ (correspondingly, $N+4) $ Weyl fermions in the antifundamental representation. A well-known example of this
type is the SU(5) gauge theory with one chiral fermion in each of two representations:
one ${\bf 10}$ and one $\bf {\bar 5}$.

We also propose  a new method of studying dynamics of the  
chiral gauge theories on 
$R_4$ by using the concept of large-$N$ 
volume independence, or the Eguchi--Kawai (EK) 
reduction.  Dynamics of any asymptotically free 
confining gauge theory formulated on $R_{4-d} \times  T_d$ 
in the $N=\infty$ limit is independent of the   size of the $d$-torus $T_d$, 
as long as the center symmetry is unbroken  
\cite{Eguchi:1982nm, Yaffe:1981vf, Bhanot:1982sh}.  
Our deformation of the chiral gauge theories indeed stabilizes 
the center symmetry  
in the small-$S_1$ regime.  
Hence, in the $N=\infty$ limit, our suggestion 
provides a  fully reduced model  for strongly coupled chiral gauge theories. 
  
Reconciliation of the  volume independence in the $N=\infty$ limit, (which is the  same as the absence of a weak coupling long distance description) and existence of a  semiclassically tractable  
small-$r(S_1)$ 
domain is also non-trivial. The domain of validity of our long-distance 
description is given by
\begin{equation}
L N \Lambda \ll 1\,.
\end{equation}
Here  
 \beq
 L =  2\pi\, r(S_1)
 \label{lrs1}
 \eeq
and $\Lambda$ is the strong scale.   
As $N \rightarrow \infty$, the  region of validity 
of our analysis  shrinks  to zero in a correlated manner, 
in accordance with the large-$N$ volume independence.   
  
\section{Chiral orbifold gauge theories: generalities}
\label{cogtg}

\begin{figure}
\begin{center}
\includegraphics[width=4 in]{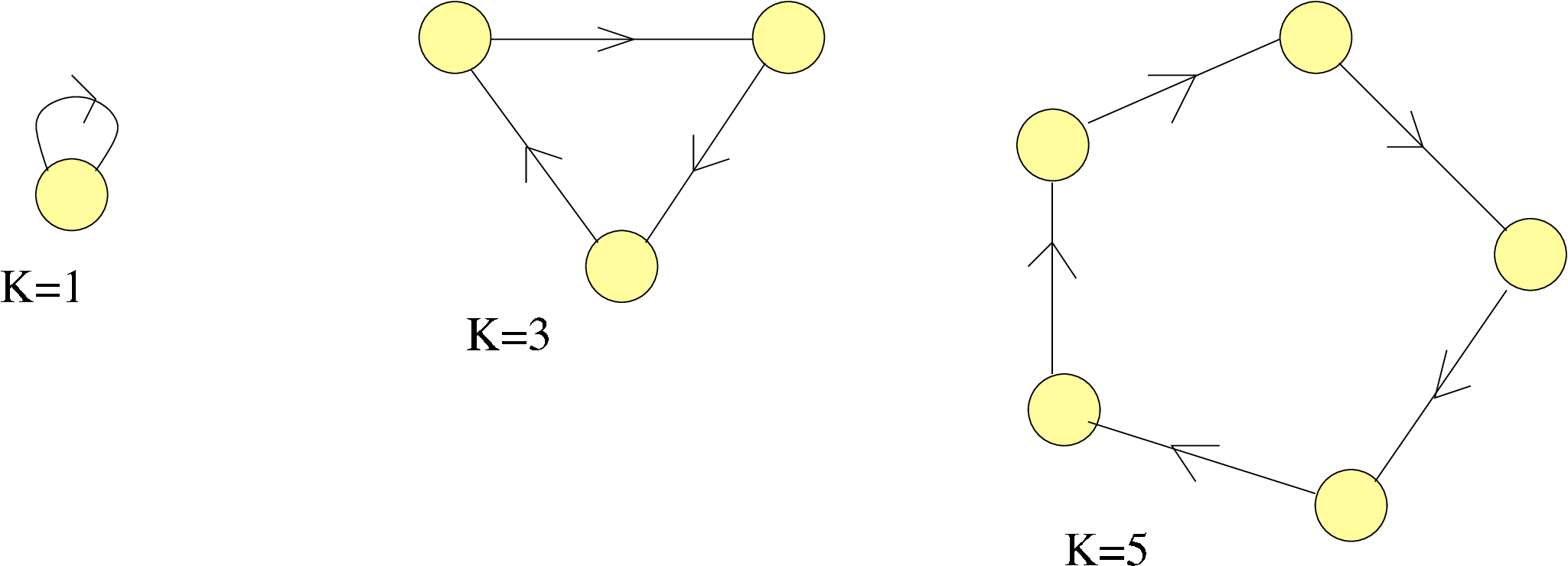}
\caption
    {\small
Orbifold gauge theories with odd number of nodes.  Nodes represents SU$(N)$ 
gauge group factors, and arrows are Weyl fermions.  
The  $K=1$ is $\N=1$ SYM theory and is vector-like. 
$K\geq 3 $ are chiral since no gauge invariant mass term can be added to
the  Lagrangian.}
\label{godd}
\end{center}
\end{figure}

Consider the orbifold  gauge theory in four dimensions  with 
the 
$$
{\rm SU}(N)_1\times  {\rm SU}(N)_2 \times \ldots \times {\rm SU}(N)_K
$$   
gauge group,
and one bifundamental  Weyl fermion on each link, 
 \begin{equation}
 \psi_J \sim (1, \ldots, N_J , \overline{N}_{J+1}, \ldots 1), \qquad  J=1, \ldots K, \; \;  
  K+1 \equiv 1\,.
 \end{equation} 
The matter content of these theories is encoded in a quiver diagram  shown in Fig.~\ref{godd}. 
The action of the theory  defined on $R_3 \times S_1$ is given by
\begin{eqnarray}
S = \sum_{J=1}^{K} \int_{R_3 \times S_1} 
\frac{1}{g^2}\,  \Tr \left[ \frac{1}{2} F_{J, MN}^2 (x)  + 
i \bar  \psi_J  \bar \sigma_M D_M \psi_J 
  \right]\,,
\label{eq:chiralorbi}
\end{eqnarray}
where the covariant derivative acts in the bifundamental representation, 
\beq
D_M \psi_J =  \partial_M \psi_J + i A_{J,M} \psi_J - i \psi_J A_{J+1, M}   \,.
\label{eq:covariant}
\eeq
The theory is chiral for $K \geq 3$. For $K\leq 2$ it is vector-like. The $ K=1$  theory is 
in fact ${\mathcal N}=1$ super-Yang--Mills (SYM), and the $K=2$ case
is the ${\rm SU}(N) \times {\rm SU}(N)$ theory with a single bifundamental 
representation fermion, which is usually referred to as  QCD(BF). 

Classically, the theory possesses $[{\rm U}(1)]^K  \times (Z_K)$ global
symmetry acting on elementary fields as
\begin{eqnarray}
&&[{\rm U}(1)]_J: \; \;   \psi_{I} \rightarrow e^{i \alpha_J \delta_{IJ} }  \psi_{I}, \;\; 
  \nonumber\\[2mm]
 &&(Z_K):  \;\; \; \;\;   \psi_J \rightarrow  \psi_{J+1}, \; \;   A_{M, J} \ \rightarrow  A_{M, J+1}  \,.
 \label{Eq:symorb}
\end{eqnarray}
where  $Z_K$ is the shift symmetry of the quiver, and  the $[{\rm U}(1)]_J$ is 
the (chiral) rotation associated with the fermion $\psi_J$.
However, quantum mechanically, the current associated 
with the chiral symmetry is not conserved.   Its  divergence  is 
  \begin{eqnarray}
\partial_{M} {\mathcal J}_{J, M} = \partial_{M} ( \bar \psi_J \bar \sigma_{M} \psi_J) = \frac{N}{16\pi^2} \left( F_J \widetilde F_J  +  F_{J+1} \widetilde F_{J+1}  \right) 
\end{eqnarray} 

For $K=2$,  which is a vector-like theory, we have 
$$
\partial_{M} ( \bar \psi_1 \bar \sigma_{M} \psi_1) =\partial_{M} ( \bar \psi_2 \bar \sigma_{M} \psi_2)\,. 
$$
Thus, the U$(1)_V$  current ${\mathcal J}_1 - {\mathcal J}_2$ is conserved. In the chiral theories with odd $K$,  there is no combination  of currents which remains conserved.  For even $K$,   the combination 
 $$ 
{\mathcal J}_M^c\equiv  \sum_{J=1}^{K} (-1)^J \, {\mathcal J}_{J,M}
$$
is conserved, $\partial_M\,{\mathcal J}_M^c=0 $. Hence a global U$(1)$ remains symmetry  of the theory for any even $K$.  

By the Atiyah--Singer index theorem, the instanton lying in the $J$-th gauge factor has $2N$ fermion 
zero mode insertions. $N$ of those come from  $\psi_{J-1}$ and the other $N$ are due to    $\psi_{J}$. 
The instanton-induced fermion vertex 
was found by 't Hooft \cite{Hooft1976}. The instanton associated with 
the gauge factor SU$(N)_J$ gives
\begin{eqnarray} 
I_{J}(x)&= &e^{-S_{J, \rm inst}}  \Big( \epsilon_{i_1, \ldots,  i_N} (\psi_{J-1})^{i_1}_{a_1} \ldots (\psi_{J-1})^{i_N}_{a_N}  \Big)
\Big( \epsilon^{k_1, \ldots,  k_N} (\psi_{J})_{k_1}^{a_1}  \ldots (\psi_{J})_{k_N}^{a_N}  \Big)  \nonumber
\\[2mm] 
&\sim& e^{-S_{J, \rm inst}}  
\det_{i, k }
  \left[  (\psi_{J-1}  \psi_{J})^{i}_{k}  \right] = e^{-S_{J, \rm inst}}  
\det  \left[  \psi_{J-1} \;  \psi_{J}  \right] \,,
\label{9}
\end{eqnarray}
where \beq
S_{J, \rm inst} =N\,S_0
\label{sinsta}
\eeq
for all $J$.
In the expression
$(\psi_{J-1}  \psi_{J})^{i}_{k} $, the contracted color indices associated with the gauge group 
SU$(N)_J$ are suppressed. The indices $(i, k)$ belong to SU$(N)_{J-1} \times$ SU$(N)_{J+1}$ 
gauge factors, the first nearest neighbors of the SU$(N)_J$ on the quiver. 
The determinant (or anti-symmetrization) produces {\em color-singlet} instanton operators. 
We would like to stress that the corresponding weight factor is exponentially small,
see Eq.~(\ref{sinsta}).

Since the fermion on the link $J$ communicates with instantons in two gauge groups 
on which it ends, instanton effects are collective.  Regardless of the value of $K$, but depending on whether it is even or odd,  the classical symmetry reduces quantum mechanically.  From Eq.~(\ref{9})
it is obvious that the symmetry is
\begin{eqnarray} 
&&
{\rm U}(1)^K \rightarrow  [Z_{2N}], \qquad \qquad  \qquad K {\rm  \; odd }\,,  
\nonumber\\[2mm] 
&&
{\rm U}(1)^K \rightarrow  [Z_{2N}] \times {\rm U}(1),  \qquad K \equiv 2m {\rm  \; even } \,.
\label{quchiral}
\end{eqnarray}
Note that the axial symmetry for all odd-$K$ theories reduces
to that of $\N=1$ SYM theory while
even-$K$ theories have the global symmetry of QCD(BF).  
 
The action of quantum symmetries on the elementary fields is as follows: 
 \begin{eqnarray}
&&
[Z_{2N}] : \; \;   \psi_I \rightarrow e^{i \frac{2 \pi\, k}{2N} }  \psi_I, \qquad k= 1,2, ..., 2N\,,
  \nonumber\\[2mm]
 &&
 (Z_K):  \;\; \; \;\;   \psi_J \rightarrow  \psi_{J+1}, \; \;   A_{M, J} \ \rightarrow  A_{M, J+1}  \,.
 \label{Eq:symorbp}
\end{eqnarray}
In the chiral quiver theories,  no gauge invariant mass term  (or gauge invariant local fermion bilinear condensate)  is possible.  

Local gauge invariant operators that are relevant and will be discussed below are 
\beq
B_J(x) =
   \epsilon_{i_1, \ldots,  i_N}  \epsilon^{k_1, \ldots,  k_N}   (\psi_{J})^{i_1}_{k_1}  \ldots 
  (\psi_{J})^{i_N}_{k_N}   \; \equiv \; \det \psi_J \,,
\label{13}
  \eeq
  and
  \beqn
R^{\rm even} (x) \!\! &=& \! \! (\psi_1)^{i_1}_{i_2}   \; (\psi_2)^{i_2}_{i_3}  \;  \ldots\;  (\psi_{2m})^{i_{2m}}_{i_1} \; 
\nonumber\\[3mm]
 &\equiv &
\tr ( \psi_1 \ldots \psi_{2m}) ,  \qquad \qquad K  \equiv 2m \,,
\label{14}
\\[4mm]
 R^{\rm odd} (x) &=& (\psi_1)^{i_1}_{i_2}   \; (\psi_2)^{i_2}_{i_3}  \;  \ldots\;  (\psi_K)^{i_K}_{i_{K+1}}   (\psi_1)^{i_{K+1}}_{i_{K+2}}  \ldots    (\psi_K)^{i_{2K}}_{i_{1}}   \nonumber\\[3mm]
 &\equiv  &
\tr ( \psi_1 \ldots \psi_K \psi_1 \ldots \psi_K) , \;  \qquad \qquad  K \; {\rm odd} \,.
\label{operators1}
\end{eqnarray}
The operator
$B_J (x)$ in Eq.~(\ref{13}) can be called {\em baryonic} and 
assigned the baryon number $B=1$. The operators $R^{\rm even} $ and $ R^{\rm odd} $
in Eqs.~(\ref{14}) and (\ref{operators1}) can be called ring and double-ring
operators, respectively.
The baryon operator is singlet with respect to the axial $[Z_{2N}]\times (-1)^B$.
Thus, it plays no role in describing the breaking patterns of the  chiral $[Z_{2N}]$ symmetry.
On the other hand, the chiral ring operators transform as 
\begin{eqnarray} 
&& [Z_{2N}] : \,\,\,   R^{\rm even}(x)    \rightarrow   e^{i\frac{2 \pi \,m}{N}}  R^{\rm even}(x) 
\,,\nonumber
\\[3mm]
&& [Z_{2N}] :  \,\,\,   R^{\rm odd}(x)    \rightarrow   e^{i\frac{2 \pi \, K}{N}}  R^{\rm odd}(x) \,.
\label{ringop}
\end{eqnarray}
Thus, they can (and will) be order parameters
determining the pattern of the chiral symmetry breaking in the chiral quiver gauge theories.
Let us define
\begin{eqnarray}
\widetilde m = m\;   {\rm mod} \; N,  \qquad  \widetilde K = K\;   {\rm mod} \; N\, ,
\end{eqnarray}
and 
\beq
\gamma (N, \widetilde m)
\eeq
denoting the greatest common divisor (gcd) of $N$ and 
$\widetilde m$.  Assuming that  the chiral condensates (\ref{ringop})
acquire a vacuum expectation value (VEV) we get the following
 chiral symmetry realizations:
  \begin{eqnarray} 
&&  
\langle R^{\rm even}\rangle \neq 0 \Longrightarrow   \left[Z_{2N}\right] \rightarrow  \left[ Z_{2 \,\gamma (N, \widetilde m)}\right] \,,
\label{chiralpat1}
\\[3mm]
&&  
\langle R^{\rm odd}\rangle \neq 0 \Longrightarrow   \left[ Z_{2N}\right] \rightarrow  \left[ Z_{2 \,\gamma
 (N, \widetilde K) }\right] \,.
\label{chiralpat2}
\end{eqnarray}
This implies occurrence of 
\begin{eqnarray}
\widetilde N= 
\left\{
\begin{array}{ll}
\frac{N}{\gamma (N, \widetilde m)} \,,\quad\mbox{even $K$} \,,\\[3mm]
\frac{N}{\gamma (N, \widetilde K)}\,, \quad\mbox{odd $K$}
\end{array}
\right.
\label{vacua}
\end{eqnarray}
isolated vacua.   Interestingly, 
if $K$ and $N$ are co-prime, the theory has a maximal chiral symmetry breaking and $N$ isolated vacua.  If $N= K$, the theory has a unique vacuum.  In general, 
depending on the relation between $N$ and $ K$, the number of vacua 
$\widetilde N$ in this class of theories 
varies between a unique vacuum and $N$ isolated vacua. The above result  disagrees with the 
statement of Ref.~\cite{stras1}  asserting the number of vacua to be $N$  regardless of the relation between $N$ and $K$. 
   
\subsection{Collective chirality}
\label{colch}
 
\begin{figure}
\begin{center}
\includegraphics[width=1.5 in]{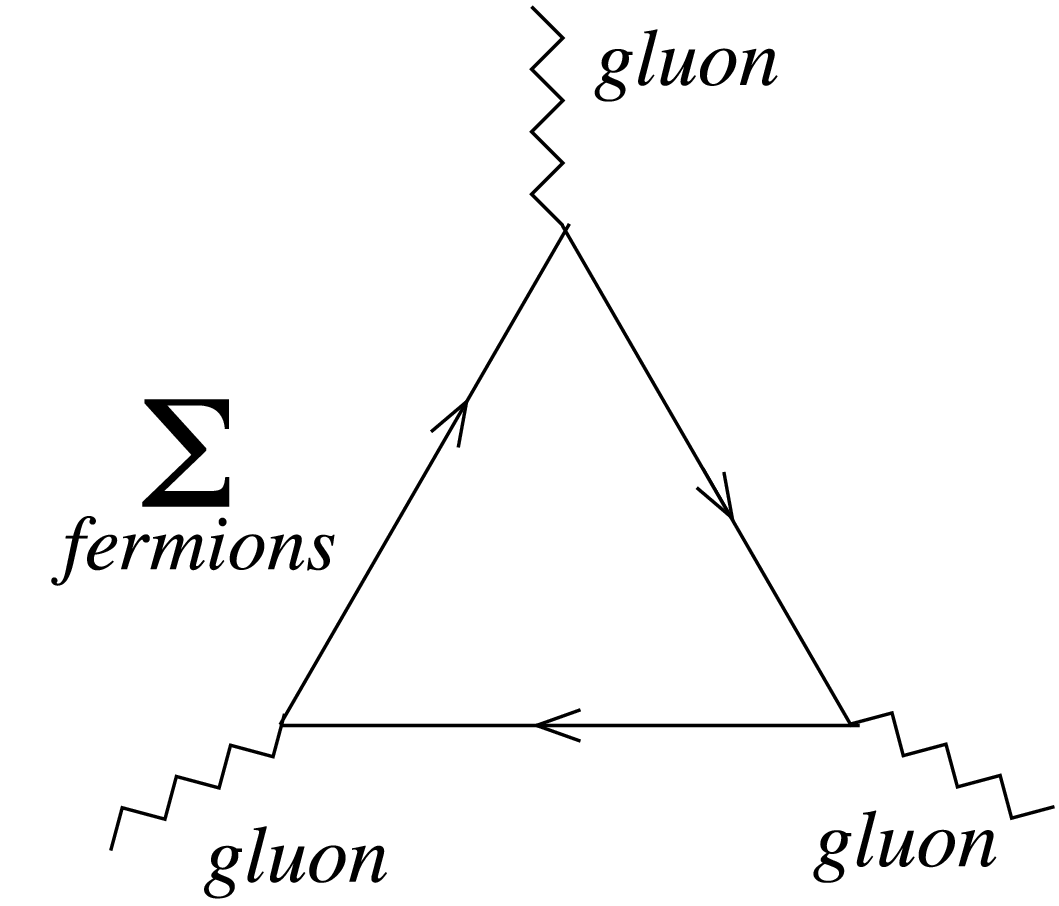}
\caption
    {\small
Internal chiral anomaly which must cancel after one sums over all fermion species in the triangle loop. }
\label{colchiral}
\end{center}
\end{figure}

 For chiral gauge theories to be consistent
 internal triangle anomalies (Fig.~\ref{colchiral}) must cancel. The textbook example is SU(5) theory
 with two Weyl fermions, one in the representations {\bf 10} and another in $\overline{\bf 5}$.
 
Unlike this old example of consistent chiral gauge theory,
in the  chiral quiver theories with $K\geq 3$ 
cancellation of the triangle anomaly proceeds in a
{\em collective manner}. 
If all gluons in Fig.~\ref{colchiral} belong to one and the same gauge  factor $[{\rm SU}(N)]_J$, 
the anomaly cancellation proceeds just like in vector-like theories
since at each $J$ we have $N$ fundamental left-handed fermions, and $N$ antifundamental.
The triangle diagram trivially vanishes if  
gluons belong to two (or three) distinct gauge factors.

One can ascribe anomaly coefficients to bifundamental fermions residing on link $J$,
 \beq
(d_{\psi_J})_{J'}=  \delta_{J J'} -  \delta_{J+1, J'}    \,.
\label{22}
\eeq
Then
\beq
(d_{\psi_{J-1}} + d_{\psi_J})_{J'}  = \delta_{J-1, J'} -  \delta_{J+1, J'}  \,.
\label{23}
\eeq
Using Eq.~(\ref{23}) we
mimic, for each $J$, 
cancellation of chiral contributions in the triangle graph of Fig.~\ref{colchiral}
inherent to vector-like theories. Note that although 
 $( d_{\psi_{J-1}} + d_{\psi_J} )_J =0$, 
the combination $( d_{\psi_{J-1}} + d_{\psi_J} )_{J \pm1} = \pm1$, exhibiting a collective chiral nature of the quiver theory  $[{\rm SU}(N)]^K$. The anomaly free nature of the latter follows from
 \beq
\sum_{J=1}^K  \; d_{\psi_J} \; = 0\,.
\eeq

It is instructive to compare
this {\em collective chirality} with a more conventional structure of 
``old" chiral gauge theories. For example, in the $SU(N)$ gauge theory with one anti-symmetric (AS) representation and  $(N-4)$ anti-fundamental $(\overline {\rm F})$ representations, we have 
$d_{AS}=N-4, \; d_{F} = +1$. Consequently, 
 \beq
d_{AS} + (N-4) d_{\overline F} = 0\,.
 \eeq

\subsection{Center symmetry and  its stabilization}
\label{srrwtdtd}

In \cite{Shifman:2008ja} and \cite{Unsal:2008ch} the center stabilizing double-trace deformations 
were applied  to the  Yang--Mills theory and QCD-like theories with a massless 
one-index and two-index representation fermion to control non-perturbative aspects 
of these  theories. In non-Abelian  vector-like gauge theories {\em without continuous 
global symmetries}, physics at small  $r(S_1)$ can be smoothly connected 
with that of the large-$r(S_1)$ theory ($R_4$ theory in the limiting case) by invoking 
such deformations.
In chiral  quiver theories we are certain that the small-$r(S_1)$  and large-$r(S_1)$ regimes will be indistinguishable by conventional order parameters within the Landau--Ginzburg--Wilson 
paradigm.\footnote{Quantum phase transitions not associated with any apparent global symmetry of the theory could occur. This possibility will be discussed separately.}
Hence, we can use the same strategy as in \cite{Shifman:2008ja}:
at small  $r(S_1)$
quasiclassical computations are possible and reliable.
{\em Raison d'\^etre} of double-trace deformations is stabilization of the
center symmetry at small  $r(S_1)$. Then qualitative lessons about 
the existence of a  mass gap, chiral condensates, chiral symmetry 
realizations, etc. will continue to hold at large  $r(S_1)$ and on $R_4$. 

Let us first discuss the center symmetry ${\mathcal G}_C$ of the chiral quiver theories.   
The gauge theory compactified on $R_3 \times S_1$ has a  global symmetry, which may be 
identified with aperiodic gauge rotations.  
In the absence of fermions, we have a decoupled ${\rm SU}(N)^K$ gauge group, with a center $[Z_N]^K$, where 
each $Z_N$ factor is the center group  of associated   ${\rm SU}(N)$ group.  
Since  the fermions are in the
bifundamental representation,  they are charged under the consecutive center group factors. 
 For example, the fermion associated with link $J$ carries charge 
\beq
Q_{\psi_J}= (0, \ldots, (+1)_J, (-1)_{J+1}, 0, \ldots, 0)  
\eeq
under the center group  $[Z_N]^K$, where each entry is defined modulo $N$.

Consider an  external object charged as 
$(q_1, \ldots, q_N) \in  [Z_N]^K$. It is easy to show that any such $N$-vector can be expressed  as 
\beq
(q_1, \ldots, q_N)  = a_0 (1, 1, \ldots , 1) + \sum_{J=1}^K a_J  Q_{\psi_J}\,.
\eeq
The external probe quarks  with any linear combinations of charges  $Q_{\psi_J}$  can be screened  by dynamical fermions. 
However, the  multiples  of $(1, 1, \ldots , 1)$ with $a_0 \neq 0\;  ( {\rm mod} \; N)$  cannot be screened by dynamical quarks, and thus, can be used as external probes  to monitor confinement.  The {\it center symmetry}  of the quiver theory  is therefore  the quotient  group 
\beq
 {\mathcal G}_C \sim \frac{[Z_N]^K}{[Z_N]^{K-1}} \sim [Z_N]_C\,.
 \label{collective}
\eeq
This is indeed the center symmetry  of pure Yang--Mills (YM) theory, $\N=1$ SYM and QCD(BF).
The presence of the complex dynamical bifundamental fermions reduces the center 
group  $[Z_N]^K$  down to the diagonal center symmetry  $[Z_N]_C$.  

It is possible to determine the realization of the $(Z_N)_C$ symmetry in the small-$S_1$ 
regime of the chiral quiver theories. 
A  one-loop potential  for the holonomies   
\beq
U_J({\bf x})=Pe^{ i \int_0^L dx_4  A_{J,4}({\bf x}, x_4)} 
 \eeq 
where $P$ denote path ordering, is induced by quantum or thermal  fluctuations.
A simple calculation within the background field  method  gives 
\beqn
V_{\rm eff}[U_J]
&=&
 \frac{2}{\pi^2 L^4} \sum_{J=1}^{K} \sum_{n=1}^{\infty} \frac{1}{n^4} \, 
   \Big[ -  |\tr \, U^n_J|^2     
\nonumber\\[3mm]
&+&
    \frac {a_n}{2} \left( \tr\, U^n_J \;  \tr \,U^{*\, n}_{J+1} + \tr\, U^{*\, n}_J \;  \tr \,U^{n}_{J+1} \right) \Big] 
   \label{oneloop}
\eeqn
where $K+1 \equiv 1 $,  $\,  a_n= (-1)^n$  for thermal (anti-periodic) spin connection 
and  $a_n= +1$  for periodic spin connection.  
The contribution of a Weyl fermion to the 
one-loop effective potential is half of the one of the 
Dirac spinor on the same background, explaining 
the appearance of the factor  $\half$ in front of the fermion 
induced terms.

The above one-loop potential also demonstrates why the global center symmetry of the theory 
is the one given in (\ref{collective}).  The gauge-boson contributions  to the potential 
(\ref{oneloop})
has a $[Z_N]^K$ symmetry  acting as 
\beq
[Z_N]_J : \; \;   \tr U_{J'}({\bf x}) \rightarrow e^{i \frac{2 \pi b_J}{N} \delta_{JJ'} } 
 \;   \tr U_{J'}({\bf x}), \qquad  
b_J=1, \ldots, N \,,
 \eeq 
while the part due to the bifundamental fermions  locks the independent phase factors $b_J$ into the  diagonal  ($J$-independent) $b$, namely,
\beq
[Z_N]_C :  \; \;  \tr U_{J}({\bf x}) \rightarrow e^{i \frac{2 \pi b}{N}  }  \;   \tr U_{J}({\bf x}), \qquad  
b =1, \ldots, N \,.
 \eeq 
This is  equivalent to the quotient construction which leaves only the diagonal $[Z_N]_C$ as the center symmetry of the theory. 

On  $S_1 \times R_3$ we can deform the original chiral theory (\ref{eq:chiralorbi}) by a center-stabilizing   double-trace deformation $P[U_J({\bf x}) ]$. The deformed action is 
\begin{equation}
S^{*} = S + \int_{R_3 \times S_1} P[U_J({\bf x}) ]
\end{equation} 
where 
 \begin{equation}
P[U_J({\bf x}) ]= A \frac{2}{\pi^2 L^4} \, \sum_{J=1}^{K} \sum_{n=1}^{\left[\frac{N}{2}\right]}  \frac{2}{n^4}         
 \left|
\, {\rm Tr}\, U^n_J ({\bf x} )\right|^2 \,,
\label{fourma}
\end{equation}
In (\ref{fourma}), 
 $A$ is an overall parameter of order one,\,\footnote{The parameter $A$ can be tuned to have 
a weak coupling center symmetry changing transition in YM*, QCD* or deformed 
chiral gauge theories.  We believe such a set-up can be useful in studying 
a non-perturbative magnetic component of the  
quark-gluon plasma, which is currently under discussion 
\cite{Chernodub:2006gu, Liao:2006ry}.  For related suggestions 
see the recent review \cite{Shuryak:2008eq}.} and ${\left[...\right]} $ denotes the integer part of 
the argument in the brackets. For sufficiently large $A$, the center  symmetry remains unbroken in the vacuum
(Sect.~\ref{peth}).
This implies a spontaneous breaking of the
non-Abelian gauge symmetry and weak coupling at small $r(S_1)$, 
which, in turn, paves the way to the
quasiclassical techniques in the chiral gauge theories. 

\subsection{Perturbation theory}
\label{peth}

We assemble together expressions in Eqs. (\ref{oneloop}) and Eqs. (\ref{fourma})
and find the stationary point.
The  center symmetry stability at weak coupling  implies that this stationary point, the vacuum of the deformed theory, is located at 
\beqn
&&
-i\,\ln\,\langle  \, U_J \rangle \equiv
L \langle \Phi_J\rangle   
\nonumber\\[4mm]
&&
= {\rm diag} \left( -\frac{2\pi [N/2]}{N},\,\,  -\frac{2\pi ([N/2]-1)}{N}, ....,\,
\frac{2\pi [N/2]}{N} \right) 
 \label{Eq:pattern}
\eeqn
modulo $2\pi$.
Consequently, in the weak coupling regime, the gauge symmetry is broken,
\begin{equation}  
[{\rm SU}(N)]^K  \longrightarrow [{\rm U}(1)^{N-1}]^K
\label{gaugesym}
\end{equation}
to a rank-preserving maximal Abelian subgroup.\footnote{In chiral gauge theories on  $R_4$, 
there are two hypotheses on dynamics of the theory \cite{Dimopoulos:1980hn}, see also 
the lectures \cite{Peskin:1982mu}.
One is the  Higgs picture  in which gauge symmetry breaks 
itself spontaneously (to a rank reducing non-abelian subgroup) and another is symmetric confinement picture.  Dimopoulos, 
Raby and Susskind introduced the idea of complementarity of these two descriptions. 
Our approach to chiral dynamics is reminiscent of the idea of complementarity, but different. 
In our case,  the small-$r(S_1)$ regime presents a  Higgs regime (albeit down to a rank preserving abelian subgroup), but the theory exhibits 
Abelian confinement due to non-perturbative effects. At 
large $r(S_1)$ (strong coupling),  we expect a  non-Abelian
confinement. (See Fig.\ref{saturation} and the discussion in section \ref{ralnl}.) 
One difference of our small-$r(S_1)$ Higgs regime 
and the Higgs picture  of~\cite{Dimopoulos:1980hn} is that in the former, 
the rank is preserved and in the latter, the rank is reduced.} 
In perturbation theory $K(N-1)$  photons remain massless while
all off-diagonal gauge fields, ``$W$ bosons,"  
acquire masses in the range $\left[\frac{2 \pi}{LN},    
\frac{2 \pi}{L} \right]$. 
The diagonal components of the bifundamental Weyl fermions 
$$
(\psi_J)^i_i \equiv \psi_{J,i}, \qquad  J=1, \ldots, K,  \; \; i=1, ..., N
$$
remain massless to all orders in perturbation theory. 
The off-diagonal fermions and $W$ bosons acquire masses 
$$m_{(\psi_J)^{i}_{k}} =m_{(W_J)^{i}_{k}} =  2\pi \frac{  (i-k) \;  {\rm mod}\;  N }{L}, $$ and decouple in the low-energy limit. Similarly, the fluctuation of the eigenvalues acquire masses proportional to $g/L$, and are also unimportant at large distances. 
The eigenvalue distributions of  
$\ln U_J$  are essentially pinned at the bottom  
(\ref{Eq:pattern}) of the 
combined  $V_{\rm eff}[U_J]+ P[U_J]$  potential. 

The electric charges of each bifundamental fermion $\psi_J$ are 
characterized by concatenation of   two $(N-1)$-dimensional vectors
 under $$ [{\rm U}(1)^{N-1}]_J \times  [{\rm U}(1)^{N-1}]_{J+1}$$
 and neutral under other gauge group factors. Namely,  
\beq
\mbox{\boldmath $q$}_{\psi_{J, i}} = g\, (0, \ldots, 0,  \underbrace{+ \mbox{\boldmath $H$}_{ii}}_{J}, \underbrace{ - \mbox{\boldmath $H$}_{ii}}_{J+1},  0, \ldots, 0
 ) \;  \in \;   [{\rm U}(1)^{N-1}]^K
 \eeq
 where  
 \beq 
  g  \mbox{\boldmath $H$}_{ii}\equiv g (H_{ii}^1, \ldots, H_{ii}^{N-1} )  \in [{\rm U}(1)^{N-1}]_J
 \eeq
are the electric charges coupled to the  $(N-1)$ photons $A_{J,\mu}^a$ 
($a=1, \ldots , N-1$), and $\mbox{\boldmath $H$}$ are the Cartan generators.

 If we define 
\beq
 \Psi_J \equiv \left[ \begin{array}{ccc}
  \psi_{J,1}  &  &  \\
  & \ddots & \\
 &  &   \psi_{J,N} 
      \end{array}
  \right], \qquad G_J \equiv \left[ \begin{array}{ccc}
  e^{i \xi_{J,1} }  &  &  \\
  & \ddots & \\
 &  &  e^{i \xi_{J,N} }    
      \end{array}
  \right], 
\eeq
 where  $e^{i \xi_{J,i} } \equiv   e^{i   \mbox{\boldmath $H$}_{ii}   \mbox{\boldmath $\xi$}_J }$, then  the gauge invariance of the low-energy theory  takes the form
\begin{eqnarray}
 \Psi_J \rightarrow G_J  \Psi_J  G_{J+1}^{\dagger}, \qquad  {\rm or }, \; \; \psi_{J, i} \rightarrow e^{i \xi_{J,i} }  \psi_{J, i} 
  e^{-i \xi_{J+1,i} }\,.
  \label{gauge}
 \end{eqnarray} 
 Note that the gauge symmetry is not $[{\rm U}(1)^N]^K$, but, rather,   $[{\rm U}(1)^{N-1}]^K$ as it is reflected  in $K$ conditions 
 \begin{eqnarray} 
 \sum_{i=1}^{N} \xi_{J,i} (x) =  \sum_{i=1}^{N}  \mbox{\boldmath $H$}_{ii}   \mbox{\boldmath $\xi$}_J(x) =0\; ,
 \end{eqnarray}
which follow from   $\sum_{i=1}^{N}  \mbox{\boldmath $H$}_{ii} =0$.
Thus, the low-energy effective  Lagrangian in perturbation theory is 
 \begin{eqnarray}
 &&
{\mathcal L}^{\rm pert\,\, th}= \sum_{J=1}^{K}  \;  \frac{1}{g_3^2}\,  \Big[ \sum_{a=1}^{N-1}\,
\Big( \frac{1}{4} F^{a}_{J, \mu \nu} \Big)^2
\nonumber\\[3mm]
 &&+ 
    \sum_{i=1}^{N}   i \bar \psi_{J,i} \gamma_{\mu} \Big( \partial_{\mu}   + i
     \mbox{\boldmath $H$}_{ii} \mbox{\boldmath $A$}_{J, \mu}  - i
      \mbox{\boldmath $H$}_{ii} \mbox{\boldmath $A$}_{J+1,\mu}   
 \Big)  \psi_{J,i}
    \Big]   \,.
    \label{pert}
\end{eqnarray}
This is an all-orders result in perturbation theory.  
Some erroneous results that one could deduce from the perturbative analysis are
(i)  the  absence of the mass gap for 
$[{\rm U}(1)^{N-1}]^K$ gauge fields (photons); (ii) enhancement of the 
discrete $(Z_N)_A$ symmetry to continuous $[{\rm U}(1)]^N$ 
fermion number symmetry 
(this is the fermion number symmetry from the standpoint of 
three-dimensional large-distance physics); 
and (iii) no chiral symmetry breaking and no stable flux tubes. 

However, all the above conclusions of the perturbative analysis are incorrect, 
as it was the case also in the nonchiral theories~\cite{Shifman:2008ja}.  The most interesting features, such as a mass gap in the gauge sector, stable flux tubes, discrete chiral symmetry breaking are 
induced due to non-perturbative effects which are invisible in the perturbative analysis. 

All non-perturbative effects listed above arise due to non-perturbative dynamics
in the quasiclassical approximation.  We 
will identify and classify non-perturbative effects induced by topologically
nontrivial field configurations momentarily.   By topologically non-trivial configurations, 
we do not mean only monopole-instantons (fractional instantons)  or instantons.  In fact, the former play no role  in chiral gauge theories (as opposed to the vector-like theories where they play a key role). Interestingly, there exists a new class of topological excitations. 
Below we give a brief  introduction to such  {\em flux operators}. 
After a general characterization, we will return to 
non-perturbative description of the chiral quiver gauge theories. 

\section{Flux operators or   instanton-monopole mo\-lecules}
\label{foaimm}

In non-Abelian gauge theories  in which gauge symmetry reduces to 
maximal Abelian subgroup at large distances, as in  (\ref{pat1}) or (\ref{pat2}),   
there are generically stable topological excitations. 
These excitations are naturally described in 
framework of the $e^{-S_0}$ expansion where 
$S_0$ is the action of the corresponding quasiclassical field configuration.   

Below, we show that the non-perturbative dynamics of the chiral theories  
on  small $S_1 \times R_3$  are quite exotic, and 
very different from the deformed YM theory~\cite{Unsal:2008ch} 
  and vector-like QCD* theories \cite{Shifman:2008ja} in the same regime. 

Perhaps, the most interesting of all is the vanishing of the monopole-instanton  
operators in  
the chiral orbifold theories and in chiral gauge theories in general!  
This excludes the so-called ``monopole mechanism"
of confinement in the chiral theories. Despite the absence of the monopole operators, there are 
other magnetically charged  excitations and flux-carrying operators.  

The flux operators carry either magnetic or topological charges, or both.  In the quiver 
gauge theories formulated on $R_3 \times S_1$ these charges are 
\begin{equation}
\left( \int_{S^2} F_J,  \qquad  \int_{R_3 \times S_1}  
\frac{g^2}{32 \pi^2} F_{J}^a  {\widetilde F}_{J}^{a} \right)\,.
\label{topcharges}
\end{equation}
Any excitation for which either of these two charges does not vanish is either an elementary 
or composite topological excitation. They can be  classified according to the powers of
 \beq
  e^{- S_0} \equiv e^{- \frac{8 \pi^2}{g^2N}} \,.
  \eeq
In a typical (quiver) gauge theory at small $r(S_1)$,  some relevant  flux operators are   
\\
\mbox{}
 \begin{eqnarray} 
&&
  { \rm monopoles:}\qquad   e^{-S_0} e^{+i \mbox{\boldmath $\alpha$}_i\,  \mbox{\boldmath $\sigma$}_J} \; {\mathcal O}_1(\psi)\,,   \nonumber \\[3mm]
&&
{ \rm bions:}\qquad   e^{-2S_0}  e^{+i (\mbox{\boldmath $\alpha$}_i -\mbox{\boldmath $\alpha$}_{i\pm 1})\;     \mbox{\boldmath $\sigma$}_J } \,,   \nonumber \\[3mm]
 && 
{ \rm BPST{\mbox{-}}instantons:} \qquad  e^{-NS_0}  {\mathcal O}_2(\psi)\,,    \nonumber \\[4mm]
&&
{ \rm flux \;  (monopole) \; rings:} \qquad  e^{-KS_0}  e^{+i \mbox{\boldmath $\alpha$}_i \sum_{J}   \mbox{\boldmath $\sigma$}_J }   {\mathcal O}_3(\psi) \,,
\label{44}
 \end{eqnarray}
where ${\mathcal O}_{1,2,3}$ are various fermionic structures:
${\mathcal O}_{2}$ can be read off from Eq.~(\ref{9}),
${\mathcal O}_{1}$ and ${\mathcal O}_{3}$ are presented in Eqs.~(\ref{try1})
and (\ref{mreven}), (\ref{mrodd}), (\ref{fluxringo}),  
respectively.\footnote{The monopole contribution
vanishes upon integration over the U(1) collective coordinates, see below.}
A sharp distinction between the above field configurations are in their magnetic and topological quantum numbers (\ref{topcharges}) whose examples are   
   \begin{eqnarray} 
&&
{ \rm monopoles:}\qquad  \left( \pm \frac{4 \pi}{g_3}   \mbox{\boldmath $\alpha$}_i \; , \;   \pm \frac{1}{N} \right) \,,\nonumber
 \\[3mm]
&&
{ \rm bions:}\qquad  
  \left( \pm \frac{4 \pi}{g_3} 
 \left(\mbox{\boldmath $\alpha$}_i -\mbox{\boldmath $\alpha$}_{i\pm 1}\right)  \; , \;    0 \right) \,,
  \nonumber \\[3mm]  
&& { \rm BPST{\mbox{-}}instantons:} \qquad  
 \left( 0  \; , \;    \pm 1 \right) \,,
   \nonumber \\[3mm]
&&{ \rm monopole \; rings:} \qquad  
\left(  \pm \frac{4 \pi}{g_3}   ( \mbox{\boldmath $\alpha$}_{i_1}, \ldots,   \mbox{\boldmath $\alpha$}_{i_K})  \; , \;  \pm ( \frac{1}{N}, \ldots,  \frac{1}{N} ) \right) \,,
\label{45}
 \end{eqnarray}
where the first number in the parentheses stands for the magnetic number,
while the second for the topological number.
 
The  monopole operators  is a subclass of the flux operators. In a theory without massless 
fermions, ${\mathcal O}_1(\psi)$ reduces to unity. In  theories with massless fermions,  the monopole 
operators carry a certain number of fermionic 
zero modes depending on its Callias index \cite{Callias}.  This determines the form of the ${\mathcal O}_1(\psi)$ operator. 

The BPST-intantons \cite{BPST}
carry no magnetic flux, just a net topological charge. 
The instanton generates certain number of the fermion zero modes dictated by 
the Atiyah--Singer index theorem.  
At small $r(S_1)$ the Callias index theorem carries more refined data than the 
 Atiyah--Singer theorem.  As was throughly discussed  in \cite{Shifman:2008ja},  in a 
 certain sense the BPST instanton can be viewed as a composite of $N$ types of elementary monopoles.  
The sum of the Callias indices of ``constituent" monopoles is equal to the Atiyah--Singer index, and the product of the monopole operators produces the BPST-instanton vertex. 
Both topological excitations are well known. 

On the other hand, the existence and role of the magnetic bions --
topological excitations which carry a magnetic flux, but have no Callias index, 
and hence, no fermion zero modes -- was realized quite recently.  They are 
responsible for the mass gap and confinement in a large class of the  
QCD-like gauge theories  at small $r(S_1)$ \cite{Shifman:2008ja}. 
We will show that  the magnetic bions also appear in the chiral gauge theories, 
and generate a mass gap in the gauge sector.

In the chiral gauge theories there is a new and very interesting class of flux 
operators carrying both the magnetic flux and fermion zero mode insertions. 
In fact, they determine the chiral symmetry realization.  
The structure of these operators is rather unique and special to 
the chiral theory of interest. 
We will refer to them as {\em monopole ring} operators.  They are not 
limited to the quiver gauge theories and exist virtually in any chiral gauge 
theory, for instance, in those discussed in Sect.~\ref{ctotst}. 
The dynamical role of the monopole ring operators in the issue of the chiral 
symmetry is similar to that of  monopoles (``fractional instantons") in  
$\N=1$  SYM theory and QCD(AS/S/BF)* theories. 

It is desirable to give a fuller classification of the flux operators in both vector-like and 
chiral gauge theories. Here, we introduce only the flux operators which 
will capture the leading non-perturbative 
physics of these theories in the $ e^{-S_0} $ expansion. 

Below, we will discuss non-perturbative dynamics of the chiral quiver theories. Since there are some noteworthy differences between the  $K$-even and $K$-odd  cases, we examine  them separately. 
  
\subsection{$K$-even chiral orbifolds ($K=2m$)}
\label{keco}

In the chiral quiver gauge theories, there are very severe restrictions on the 
form of the flux operators. To explain the point, let us start from the 
simplest case, a decoupled pure 
 $[{\rm SU}(N)]^K $ gauge theory at small $r(S_1)$.   The presence of 
 the double-trace deformations  leads to the gauge symmetry 
 breaking,  $[{\rm SU}(N)]^K \to [{\rm U}(1)^{N-1}]^K$. In this theory a set of disentangled 
 monopole operators emerges,
\begin{eqnarray}
e^{+i \mbox{\boldmath $\alpha$}_i\,  \mbox{\boldmath $\sigma$}_J}, \qquad i=1, \ldots N, 
\qquad J=1, \ldots K \,,
\end{eqnarray}
where   $\mbox{\boldmath $\sigma$}_J$ is the dual photon
associated with gauge group [U(1)$^{N-1}]_J$, and $\frac{4 \pi}{g} \mbox{\boldmath $\alpha$}_i $ 
is the magnetic charge of the  monopole. 
Each monopole has four 
bosonic zero modes, three associated with the center 
of mass position of the monopole ${\bf x} \in R_3$ and a U(1) 
angle associated with the global part of the gauge rotations.
 
 In the presence of massless bifundamental Weyl  fermions, 
there must be  two  fermion zero modes associated with each monopole. Naively, one expects 
\begin{eqnarray}
{\mathcal M}_{J,i}  = e^{-S_0} e^{+i \mbox{\boldmath $\alpha$}_i\,  \mbox{\boldmath $\sigma$}_J } \left({ \psi_{J-1,i} \;  \psi_{J,i} 
+  \psi_{J-1,i+1} \;  \psi_{J, i+1}   } \right) \,,
\label{try1}
\end{eqnarray}
as a consequence of the  Callias index theorem.    From the point of view of the $J$-th
gauge group factor, there is nothing wrong with this operator.   However, 
in the chiral quiver theories,  the monopole operator at the quiver site $J$ also transforms non-trivially under the global gauge rotations of the nearest-neighbor  ($J\pm 1$) sites.  For example, 
\begin{eqnarray}
 \left( \psi_{J-1,i} \psi_{J,i} \right) \rightarrow e^{i \phi_{J-1, i} }
 \left( \psi_{J-1,i} \psi_{J,i} \right) e^{-i \phi_{J+1,i} } \,,
 \label{global1}
\end{eqnarray}
where  $e^{i \phi_{J\pm1, i} }$ is a global phase. This is the global part of the gauge rotations 
$e^{i \xi_{J,i}({\bf x}) } $ given in Eq.~(\ref{gauge}). 
In order to construct a manifestly global-rotation-invariant monopole operator, we have to average over all distinct U$(1)$ angles. The integral  $\int  d\phi_{J,i} \prod_{J' \neq J \pm1} 
d\phi_{J', i}$ is trivial and produces only an overall numerical factor. 
For the quiver gauge theories with $K \geq 3$ averaging over the global zero modes of the monopole on the nearest neighbor quiver sites yields 
\begin{eqnarray}
\!\!
\int d\phi_{J-1,i}  d\phi_{J+1,i}  \; 
 e^{i \phi_{J-1,i}}
  \Big[ e^{-S_0}  e^{+i \mbox{\boldmath $\alpha$}_i\,  \mbox{\boldmath $\sigma$}_J } 
  \left( \psi_{J-1,i} \psi_{J,i} \right) \Big]_{\phi_{J\pm1}=0} 
  e^{-i \phi_{J+1,i}}
 \; =\;  0
 \label{mon0}
\end{eqnarray}
due to either  of the integrations,  $\int  d\phi_{J-1,i}$ or $\int d\phi_{J+1,i}$. 
Thus, the monopole operators vanish!

Note, however, that the above integral does not vanish for $K=1$, which is $\N=1$ SYM theory and for $K=2$ which is QCD(BF)*. Both of these theories are vector-like, and the monopole contribution to the dynamics is non-vanishing as was observed previously 
\cite{Davies:1999uw, Shifman:2008ja}. 

Let us reiterate our striking conclusion:  {\em the monopole  operator
contributions   in 
non-perturbative dynamics of the chiral quiver gauge 
theories vanish identically.}  Upon averaging over all zero modes, in particular, the global U$(1)$ angles,   all monopole operators drop out. 

The structure of the monopole operators and transformation properties under the global rotations 
given in Eq.~(\ref{global1}) also suggest how to construct flux operators invariant under the 
the global U$(1)$ symmetries of the theory. If we take the product of the ``naive" $K/2=m$ monopole operators (see Eq.~(\ref{try1})) separated by 
two units in the quiver diagram, the resulting topological excitation 
will present a gauge invariant flux operator with $K=2m$ fermion insertions.   
There are two types of the monopole ring operators associated with product of the naive monopole operators on even and odd sublattice of the quiver. 
For even quiver sublattice these flux operators are
\begin{eqnarray}
{\mathcal M R}_i^{\rm even} (x)= &&e^{-\frac{K S_0}{2}}
\Big( \prod_{J \in {\rm even}} 
e^{+i \mbox{\boldmath $\alpha$}_i \mbox{\boldmath $\sigma$}_J} \Big)
  \left( \psi_{1,i} \ldots \psi_{2m,i}  \; +   \; \psi_{1,i+1} \ldots \psi_{2m,i+1}  \right)  \nonumber 
  \\[2mm] 
 \equiv &&
e^{-m S_0}
\Big( 
e^{+i \mbox{\boldmath $\alpha$}_i \sum_{J=1}^{m} \mbox{\boldmath $\sigma$}_{2J}} \Big)
 \left( R^{\rm even}_i(x) +  R^{\rm even}_{i+1}(x) \right),  
 \label{mreven}
\end{eqnarray}
while for the odd sublattice 
\begin{eqnarray}
{\mathcal M R}_i^{\rm odd} (x)= &&e^{-\frac{K S_0}{2}}
\Big( \prod_{J \in {\rm odd}} 
e^{+i \mbox{\boldmath $\alpha$}_i \mbox{\boldmath $\sigma$}_J} \Big)
  \left( \psi_{1,i} \ldots \psi_{2m,i}  \; +   \; \psi_{1,i+1} \ldots \psi_{2m,i+1}  \right) 
  \nonumber \\[2mm] 
 \equiv &&
e^{-m S_0}
\Big( 
e^{+i \mbox{\boldmath $\alpha$}_i \sum_{J=1}^{m} \mbox{\boldmath $\sigma$}_{2J-1}} \Big)
 \left( R^{\rm even}_i(x) +  R^{\rm even}_{i+1}(x) \right),  
 \label{mrodd}
\end{eqnarray}
where $i\in [1,..., N]$. 

The fermionic structure of the  two flux operators,
 $$ {\mathcal M R}_i^{\rm even} (x)\,\,\, {\rm and}\,\,\, 
{\mathcal M R}_i^{\rm odd} (x)\,,$$
is identical. 
Clearly, the monopole ring operators are not forbidden
by symmetries of the theory  and are consistent with natural generalization of the 
Callias index theorem. Namely, these operators  have $K/2=m$ constituent monopoles, 
and $K=2m$ fermion zero mode insertions.  However, as was noted above, the notion of 
a constituent monopole-instanton is somewhat misleading since such ``constituents"
do not exist in the isolated state, nor do they contribute to non-perturbative dynamics. 
In addition to 
Eqs.~(\ref{mreven}) and (\ref{mrodd}), certainly, 
there are conjugates of these topological excitations, to be labeled as $\overline {\mathcal M R}^{\rm odd/even}_i (x)$.
 
We will see that the  flux ring operators are responsible for  various non-perturbative phenomena, such as the discrete chiral symmetry realization.  In particular, note that the chiral order parameter defined in (\ref{operators1}) is related to the fermion zero mode structure of the flux operators as follows:
\begin{equation}
R^{\rm even} (x) = \sum_{i=1}^{N} R^{\rm even}_i (x) + \;  (\rm massive\;   modes).
\end{equation}
Exotic chiral condensates can be saturated by the flux ring operators, much in the same way
as the usual chiral condensates are saturated by the monopole operators at small $r(S_1)$ in $\N=1$ SYM theory and QCD(BF)* \cite{Davies:1999uw, Shifman:2008ja}.  
Just like in the $\N=1$ theory and QCD(BF)*, 
the monopole ring  operators with the fermion zero mode  insertions have nothing to say on the issue of the mass  gap and confinement in the chiral gauge 
theory~\cite{Shifman:2008ja}.  

\subsubsection{Mass gap in the gauge sector}
\label{mggs}

 In perturbation theory,  a mass term for photon is forbidden. Thus, we are searching  for 
non-perturbative effects that may generate a mass gap in the gauge 
sector of the theory. Let us first show that mass gaps for the photons 
are allowed by the symmetries of the 
microscopic theory. 

Since the symmetry 
of the microscopic theory  is ${\rm U}(1)_V \times [(Z_{2N})] \times Z_K$,  it must be manifest in  the  low-energy effective theory. 
The invariance of the monopole ring operators (\ref{mreven}) and (\ref{mrodd}) 
under the $ [(Z_{2N})]$ discrete chiral symmetry 
requires intertwining of  the axial chiral symmetry  with a  discrete shift symmetry of the 
dual photons, 
 \beqn
[(Z_{2N})]: \; \;\;\;  R_i(x) && \longrightarrow e^{i \frac{2 \pi  \widetilde m }{N}}  R_i(x)
, \nonumber\\[3mm]
 \sum_{J \in {\rm even}} \mbox{\boldmath $\sigma$}_{J}  && \longrightarrow   \sum_{J \in {\rm even}}
 \mbox{\boldmath $\sigma$}_{J}
 -  \frac{2 \pi \widetilde m}{N} \mbox{\boldmath $\rho$}  \,,
 \label{Eq:symorb2}
\eeqn
where $ \mbox{\boldmath $\rho$} $ is the Weyl vector defined by 
\beq
\mbox{\boldmath $\rho$} =  \sum_{k=1}^{N-1} \mbox{\boldmath $\mu$}_k\,, 
\label{dop1}
\eeq
and \mbox{\boldmath $\mu$}$_k$'s stand for the $N-1$ fundamental weights 
of the associated Lie algebra, defined through the  reciprocity relation, 
\beq
\frac{2 \mbox{\boldmath $\alpha$}_i \mbox{\boldmath $\mu$}_j }
{ \mbox{\boldmath $\alpha$}_i^{2}}=   \mbox{\boldmath $\alpha$}_i \mbox{\boldmath $\mu$}_j     = \delta_{ij}\,.
\label{dop2}
\eeq
Using the identities 
\begin{equation}
 \mbox{\boldmath $\alpha$}_N   \mbox{\boldmath $\rho$} =  -(N-1) \,,  \quad \mbox{\boldmath $\alpha$}_i  \mbox{\boldmath $\rho$}= 1\,
  , \quad  i=1,\,  \ldots\,  N-1\; , 
\label{iden}
\end{equation}
we see that the flux operator  
\begin{equation}
\left( \prod_{J \in {\rm even}} e^{+i \mbox{\boldmath $\alpha$}_i\, \mbox{\boldmath $\sigma$}_J}
 \right) 
\rightarrow 
e^{-i \frac{2 \pi \widetilde m }{N} } \;\left( \prod_{J \in {\rm even} } e^{+i \mbox{\boldmath $\alpha$}_i\, \mbox{\boldmath $\sigma$}_J }
 \right) 
 \,, \quad i=1,\, \ldots, \, N\,,
\end{equation}
i.e. rotates in the opposite direction compared to the $2m$-linear fermion  
ring operators $R_i(x) $, by the same amount. Hence, the monopole ring vertex  (\ref{mreven})    
is invariant under the discrete  $ [(Z_{2N})] $ chiral symmetry.  Note that the discrete shift 
symmetry  acting on the dual photons is 
 \begin{equation}
 Z_{ \widetilde N}, \qquad  {\rm where } \; \;  \widetilde N =  
  \frac {N}{{\gamma}({\widetilde m}, N)}     \; .
 \end{equation}
Recall that $\gamma({\widetilde m}, N) ={\rm gcd}({\widetilde m}, N)$.
This discrete shift symmetry, as opposed to the continuous shift symmetries, 
cannot prohibit mass terms for scalars; at best it can defer the appearance of a mass term 
to higher levels of the 
$e^{-S_0}$ expansion.  Thus, the scalar mass terms will indeed be generated. 
The flux operators such as $e^{-S_0} e^{+i \mbox{\boldmath $\alpha$}_i\,  \mbox{\boldmath $\sigma$}_J}$ are  
forbidden by the shift symmetry and are not allowed by the index theorem. There are topologically 
null, but magnetically charged excitations in the theory referred to as the magnetic bions. 
The magnetic  bion operators are
\begin{eqnarray}
e^{-2S_0} e^{+i (\mbox{\boldmath $\alpha$}_i     - \mbox{\boldmath $\alpha$}_{i\pm1}  )
\,  \mbox{\boldmath $\sigma$}_J} \,,
\end{eqnarray}
which is roughly the product of the monopole and anti-monopole operators, stripped off their  fermionic modes. 
The magnetic bion contribution to the non-perturbative part of the Lagrangian is 
 \begin{eqnarray}
\Delta {\mathcal L}_{\rm bions} =
e^{-2S_0} \sum_{J=1}^{2m} \sum_{i=1}^{N} 
 \cos {(\mbox{\boldmath $\alpha$}_i     - \mbox{\boldmath $\alpha$}_{i-1}  )
\,  \mbox{\boldmath $\sigma$}_J} 
\,.
\label{d62}
\end{eqnarray}
This is sufficient to render all photons massive.  Defining the Fourier transform of the 
dual photons as
\begin{eqnarray}
{\widetilde \sigma}_{J, p} = \frac{1}{\sqrt N}
 \sum_{i'=1}^{N} e^{i \frac{2 \pi i' p }{N} }\mbox{\boldmath $H$}_{i'i'}   \mbox{\boldmath $\sigma$}_J 
  \end{eqnarray}
diagonalizes the mass matrix leading to the following masses for the dual photons: 
\begin{eqnarray}
m_{{\widetilde \sigma}_{J, p} } &=& e^{-S_0} \left(2 \sin \frac{\pi p}{N} 
\right)^2 \nonumber  \\[4mm]
& =& \Lambda (\Lambda L N)^2  \left(2 \sin \frac{\pi p}{N} \right)^2 ,  \qquad 
J \in [1,K], \; \; p \in [1, N-1] \,.
\label{photonmass}
\end{eqnarray}  
In the second line we restored dimensions and used the one-loop 
renormalization group relation  $(\Lambda LN)^{3}= e^{-\frac{8 \pi^2}{g^2N}}$. 
The $J$ independence of this mass formula is an artifact of our truncation of
the $e^{-S_0}$ expansion at the leading order,  $e^{-2S_0}$.
The $J$-degeneracy  will be lifted  by subleading terms in the $e^{-S_0}$ expansion.  
For our purposes it is most important that all $(N-1)K$ dual photons 
of the chiral theory acquire masses at this order. 

The effect due to the operators (\ref{44})  in the large-distance effective Lagrangian is
\begin{eqnarray}
{\mathcal L}^{\rm effective} &= &  {\mathcal L}^{\rm pert \; th} +   {\mathcal L}^{\rm non \; pert}    \nonumber 
  \\[2mm] 
&=& {\mathcal L}^{\rm pert \; th} +   \Delta{\mathcal L}_{\rm bions} + \Delta{\mathcal L}_{\rm monopole \; ring }  +  
 \Delta{\mathcal L}_{\rm instantons } + \ldots 
 \label{longdis}
\end{eqnarray} 
The physics that this Lagrangian encapsulates is the main result of our work. 
The  flux operators in the large-distance  Lagrangian  
(\ref{longdis}) present microscopic sources for various non-perturbative phenomena. The dual photon masses are generated by {\em magnetic bions}. 
Linear confinement ensues much in the same way
as in non-chiral theories.
The chiral condensates are saturated by 
the {\em monopole ring operators}. Below, we will discuss the chiral condensates in 
the quiver theories in more detail. 

\subsubsection{Chiral condensates and chiral  symmetry realization}
\label{chiral}

The chiral  condensate $\langle \tr (\psi_1 \ldots \psi_{2m}) \rangle$ 
is dominated by a contribution from the flux ring operators (\ref{mreven})  and  
(\ref{mrodd}).  The chiral condensate operator has $2m$ fermion insertions. It is saturated by 
the  zero mode structure of the flux ring operators. 
This is analogous to $\N=1$ SYM theory at small $r(S_1) $  where the chiral condensate is saturated by the monopole operators with two zero mode insertions \cite{Davies:1999uw}.  
The chiral condensate  is proportional to 
\begin{eqnarray}
\langle \tr (\psi_1 \ldots \psi_{2m}) \rangle &\sim&  \sum_{i=1}^N \; \; 
\langle \tr (\psi_1 \ldots \psi_{2m}) \rangle_{ {\mathcal M R}_i^{\rm even/odd}} \nonumber \\ 
&=& 2N e^{-mS_0}e^{i \frac{2 \pi m }{N}} \,.
\end{eqnarray}
Expressing it  in terms of the
strong scale by using the one-loop result for the $\beta$ function,  we obtain  
\begin{eqnarray}
\langle \Omega_q | \tr (\psi_1 \ldots \psi_{2m} ) | \Omega_q \rangle = 
2N \Lambda^{3m} e^{i \frac{2 \pi q}{\widetilde N}}, \qquad  q=1, \ldots, {\widetilde  N}\,.
\label{condensate}
\end{eqnarray}
This shows that the chiral symmetry breaking pattern of the theory is the one given in 
Eq.~(\ref{chiralpat1}).  As was anticipated in  (\ref{vacua}),  the theory possesses  
${\widetilde  N}$ vacua, 
\begin{equation}
\left\{ |\Omega_1\rangle, \ldots,  |\Omega_{\widetilde N} \rangle \right\},  \qquad 
 {\widetilde  N} = 
\frac{N}
{{\gamma}(N, {\widetilde m}) }\,.
\end{equation}
The phase of the chiral condensate distinguishes these vacua. 

We can label the vacua in the 
$  \langle \tr (\psi_1 \ldots \psi_{2m}) \rangle$-plane, and study  aspects of  
 domain walls of the chiral gauge theory (in cases where there are multiple vacua). 

The chiral condensate  (\ref{condensate}) is an interesting result. It tells us that the condensate is independent of   $r(S_1)$ in the weak coupling regime.  Such radius independence  occurs in a few  
QCD-like theories as well.  These are $\N=1$ SYM and QCD(BF/AS/S)* theories with 
a single Dirac fermion.  In 
$\N=1$ SYM theory the chiral order parameter is a part of the so called chiral ring and is protected 
by supersymmetry.  In QCD(BF/AS/S)* theories the chiral condensate must coincide with that 
in $\N=1$ theory due to planar equivalence \cite{asv1,asv2,asv3, Kovtun:2005kh, UY}. Indeed, the  
microscopic quasiclassical  calculation at small $r(S_1)$ gives  the same result as the $R_4$ prediction in the framework of  planar orbifold/orientifold  equivalences.  This suggests that, perhaps, the value of the condensate in the quiver theory under consideration
remains invariant under decompactification into $R_4$.  

\subsubsection{Linear confinement}
\label{linco}

As was discussed in Sect.~\ref{srrwtdtd}, in the $[{\rm SU}(N)]^K$ chiral quiver gauge theory with dynamical 
bifundamental fermions, linear confinement (with unbreakable strings)
can be probed by external charges  with non-vanishing $N$-ality 
\begin{equation}
 k (1, 1, \ldots , 1) , \qquad  k = 1, \ldots, N-1
 \label{source}
\end{equation}
under the diagonal center group $[Z_N]_C$.  
Below we will demonstrate the existence of a linearly confining potential between 
such probe external charges.  
The corresponding tensions are determined by
the dual photon mass terms generated by the bion operator (\ref{d62}).
Precision evaluation of the tensions will not be carried out. 

The insertion of a Wilson loop $W_{{\mathcal R}}(C)$
in a representation $\mathcal R$ with non-zero $N$-ality $k$
corresponds, in the low-energy dual theory,
to the requirement that the dual scalar fields have non-trivial monodromy,
\begin{equation}
    \int_{C'}   ( d \mbox{\boldmath $\sigma$}_1, \ldots ,  d \mbox{\boldmath $\sigma$}_K)   = 2 \pi (  \mbox{\boldmath $\mu$}_k, \ldots ,  \mbox{\boldmath $\mu$}_k) \,,
\label{eq:monodromy}
\end{equation}
where $C'$ is any closed curve whose linking number  $\ell$ with $C$ is one:
\begin{equation}
{\ell}\; (C,C')=1
\end{equation}
regardless of the details of the contour  $C'$. 
In other words, in the presence of the Wilson loop $W_{{\mathcal R}}(C)$
the dual scalar fields must have a discontinuity of $2\pi \mbox{\boldmath $\mu$}_k$
across some surface $\Sigma$ which spans the loop $C$.

To evaluate a Wilson loop expectation value sourced by the charges (\ref{source}), 
one must minimize the dual magnetic bion induced action 
 in the space of field configurations
satisfying the monodromy condition (\ref{eq:monodromy}).
Adapting Polyakov's argument to our present problem, we find the string tension  
\begin{equation}
    T_k \equiv -\lim_{\mathrm {area}(\Sigma)\to\infty}
    \frac{\ln\left\langle W_{\mathcal R}(C)\right\rangle}
    {\mathrm {area}(\Sigma)}  \equiv
     \left.
    \min_{  \mbox{\boldmath $\sigma$}_J } \;
   \frac {\Delta S_{\rm bion}(   \mbox{\boldmath $\sigma$}_J  
  )}{\mathrm{area}(\R^2)}
 \right|_{\Delta (\mbox{\boldmath $\sigma$}_1, \ldots, \mbox{\boldmath $\sigma$}_K) }
\end{equation}
where
$$
 \Delta \mbox{\boldmath $\sigma$}_J = 
  \mbox{\boldmath $\sigma$}_J (\infty) -   \mbox{\boldmath $\sigma$}_J (-\infty)  =     \mbox{\boldmath $\mu$}_k 
$$ 
and $\Delta S_{\rm bion}(   \mbox{\boldmath $\sigma$}_J  )$ is the magnetic bion  action minus its vacuum value.
 Note that due to the equivalence relations 
 $$  \mbox{\boldmath $\mu$}_k 
  =  k  \mbox{\boldmath $\mu$}_1  +   \mbox{\boldmath $\alpha$}  
 $$ 
 for some   $ \mbox{\boldmath $\alpha$}  $ in the root lattice, 
  we are guaranteed to have  $T_k=T_{N-k}$  for the string tensions. 
 The $k$-string tension $T_k$ equals the mass of a kink
solution with topological charge $k$. 

 The linearly confining chiral quiver gauge theories are similar to pure YM theory 
or YM theory with adjoint fermions. In both cases, there are $N-1$ types of stable flux tubes associated with an unbroken $Z_N$ gauge symmetry in the infrared.  

The reader should also note that the monodromy condition (\ref{eq:monodromy}) is different from the change of the value of the dual photon scalar  in passing from one isolated vacuum of the theory
to another (in cases where there are multiple vacua).  The latter is associated with the spontaneous breaking of 
the $Z_{\widetilde N}$ discrete 
shift symmetry (or the discrete chiral symmetry) while the  ${\widetilde N}$ isolated vacua are separated by 
 \beqn
\Delta  \Big(\sum_{J \in \rm even} \mbox{\boldmath $\sigma$}_{J} \Big) = 
  \frac{2 \pi \widetilde m}{N} \mbox{\boldmath $\rho$} \,.
 \eeqn
 In our case, the monodromy is due to an external source probing the vacuum of the theory and  
 the jump across the interface of the Wilson loop is not associated with any spontaneous symmetry breaking.  The directions of these two types of monodromies in the field space are not parallel to each other.  

 \subsection{Odd-$K$ chiral orbifolds} 
 
Dynamics of the center stabilized   odd-$K$ chiral orbifold theories is similar  
to that of their even-$K$ counterparts. Below, we will only outline the differences. 
 
The gauge invariant flux (monopole) ring operators,  
the analogs of (\ref{mreven}) and   (\ref{mrodd}), are  
\begin{eqnarray}
{\mathcal M R}_i (x)= &&e^{-K S_0}
\Big( \prod_{J } 
e^{+i \mbox{\boldmath $\alpha$}_i \mbox{\boldmath $\sigma$}_J} \Big)
  \left( \psi_{1,i} \ldots \psi_{K,i}   \psi_{1,i} \ldots \psi_{K,i}  \; +  [ i \rightarrow i+1]
    \right)    \nonumber \\[3mm]
 \equiv 
 &&
e^{-K S_0}
\Big(  
e^{+i \mbox{\boldmath $\alpha$}_i \sum_{J=1}^{K} \mbox{\boldmath $\sigma$}_{J}} \Big)
 \left( R_i(x) +  R_{i+1}(x) \right) , \quad i=1, \ldots, N. \nonumber \\[3mm]
 \label{fluxringo}
\end{eqnarray}

As was discussed in Sect.~\ref{cogtg}, 
the microscopic theory possesses a $[(Z_{2N})]$ axial symmetry.   Hence, this must 
also be a symmetry  of the large-distance effective theory. This is possible due to   
 intertwining of the chiral symmetry with the shift symmetry of the dual photons,
 \beqn
[(Z_{2N})]: \; \;\;\;  R_i(x) && \longrightarrow e^{i \frac{2 \pi  K }{N}}  R_i(x)
, \nonumber\\[4mm]
 \sum_{J } \mbox{\boldmath $\sigma$}_{J}  && \longrightarrow   \sum_{J }
 \mbox{\boldmath $\sigma$}_{J}
 -  \frac{2 \pi K}{N} \mbox{\boldmath $\rho$}  \,.
 \label{Eq:symorb2p}
\eeqn
The $Z_{2N}$ discrete axial  symmetry transmutes into the dual photon as a discrete 
$Z_{ \widetilde N}$  symmetry where $\widetilde N$ is given in (\ref{vacua}).  The spontaneous breaking of $Z_{ \widetilde N}$ is responsible for the existence of  $\widetilde N$ isolated vacua. 
As before, if  $\widetilde N$ is equal to unity, then the $Z_{2N}$ chiral symmetry of the microscopic theory remains unbroken. 

The chiral condensate $\langle \tr (\psi_1 \ldots \psi_K \psi_1 \ldots \psi_K) \rangle$ receives its dominant contribution from the monopole ring operators  (\ref{fluxringo}) discussed above.  
Proceeding along the lines of Sect.~(\ref{chiral}), we arrive at 
\begin{eqnarray}
\langle \Omega_q | \tr (\psi_1 \ldots \psi_K \psi_1 \ldots  \psi_{K} ) | \Omega_q \rangle = 
N \Lambda^{3K} e^{i \frac{2 \pi q}{\widetilde N}}, \qquad  q=1, \ldots, {\widetilde  N}\,.
\label{condensate2}
\end{eqnarray}
This formula shows that the pattern of the chiral symmetry breaking  of the theory is that presented in 
Eq.~(\ref{chiralpat1}).  As anticipated in  (\ref{vacua}),  the theory possess  
${\widetilde  N}$ vacua, 
\begin{equation}
\left\{ |\Omega_1\rangle, \ldots,  |\Omega_{\widetilde N} \rangle \right\},  \qquad 
 {\widetilde  N} = 
\frac{N}
{{\gamma}(N, {\widetilde K}) }\,.
\end{equation}
Other aspects of the odd-$K$ chiral quiver theories are very similar to the even-$K$ case. In particular, the mechanism of the mass gap generation in the gauge sector is the
same in both cases, and is due to the magnetic bions.
 
 \vspace{5mm}
 
\section{Chiral theories of the second type}
\label{ctotst}

In this section,  we
will briefly discuss the  chiral gauge theories of a traditional type, with a single gauge group factor and a chiral matter content.  
Examples are SU$(N)$ gauge theory with one  AS  (S) Weyl fermion and $N-4$ $(N+4)$ anti-fundamental representation Weyl fermions.  
The gauge anomaly coefficient of the AS (S)  representation is $N-4$ $(N+4)$ and that of
the fundamental representation  is $+1$.  Hence, these theories are internally free
of triangle anomalies, and are self-consistent.  Below, we consider 
the theory with the AS fermions as an example. 
 
Classically, the theory possesses an  ${\rm U}(1)_a \times {\rm U}(1)_b \times {\rm SU}(N-4)$ global symmetry defined by
\begin{eqnarray}
&& {\rm U}(1)_a:  \quad \psi_{[ab]} \rightarrow e^{i \beta} \psi_{[ab]}\,,   \nonumber\\[3mm]
&& {\rm U}(1)_b: \quad \psi^{a}_{I} \rightarrow e^{i \delta} \psi^{a}_{I}, \qquad   I=1, \ldots N-4\,,
 \nonumber\\[3mm]
&& {\rm SU}(N-4):  \quad \psi^{a}_{I} \rightarrow (V  \psi^{a})_{I}, \qquad   V \in {\rm SU}(N-4)\,.
\end{eqnarray}
In the  quantum theory, due to instanton effects,
only the  ${\rm U}(1) \times {\rm SU}(N-4)$ symmetry survives. 

Recall that the BPST instanton has $N-2$  insertion of antisymmetric $\psi_{[ab]}$'s along with 
$N-4$ antifundamental fermion zero mode insertions, one for each flavor.
A manifestly  SU$(N-4)$ global symmetry invariant instanton operator is proportional to 
  \begin{equation}
I_{\rm  inst} = e^{-S_{\rm inst}} \underbrace{\psi_{[a_1b_1]} \ldots  \psi_{[a_{N-2} b_{N-2}]}}_{{N-2}}
  \underbrace{\psi^{b_1}_{I_1}  \ldots \psi^{b_{N-4}}_{I_{N-4}}}_{N-4} \epsilon^{I_1 \ldots I_{N-4}} 
   \epsilon^{a_1 \ldots a_{N-2}  b_{N-3}  b_{N-2}  } \,.
   \nonumber\\[3mm]
  \end{equation}
Clearly,  the instanton effect  spoils one particular linear combination of the
classical U$(1)$ symmetries. The linear combination which is preserved by the instanton vertex is 
 \begin{equation}
e^{i \alpha Q} \psi_{[ab]} =e^{i \alpha (N-4)} \psi_{[ab]}, \qquad e^{i \alpha Q} \psi^{a} =e^{-i \alpha 
(N-2)} \psi^{a} \,.
\label{U(1)}
\end{equation}

We consider this gauge theory at small $r(S_1)$ with either periodic or antiperiodic spin connection.  For what follows, the difference is immaterial.  The one-loop potential can be obtained as in (\ref{oneloop}),
\begin{eqnarray}
V_{\rm eff}[U_J] &&=
 \frac{2}{\pi^2 L^4}  \sum_{n=1}^{\infty} \frac{1}{n^4} \, 
   \Big\{ -  |\tr \, U^n|^2     
    \nonumber \\[3mm] 
 &&
+   \frac{ a_n }{2} \Big[
    \Big( \frac{(\tr\, U^n )^2 - (\tr\, U^{2n})}{2}  +  (N-4)  \tr\, U^n   
    \Big) +{\rm h.c.}  \Big]  \Big\} .
   \label{oneloopp}
\end{eqnarray}
where the fermionic contributions  (the second line)  is half of  the Dirac fermions in the corresponding representation.
Regardless of the spin connection, this potential exhibits attraction between the eigenvalues of the 
Polyakov line. In the thermal case, the theory will be in the deconfined high temperature phase.

At small $r(S_1)$  we add a double-trace deformation which generates a repulsion among the eigenvalues of the Polyakov line.  As opposed to the chiral quivers where there is an exact $[Z_N]_C$ center symmetry, the ``traditional" chiral SU$(N)$ theory  has no exact center symmetry. What our 
double-trace deformation does in this case, is to provides an eigenvalue repulsion rendering the vacuum  as close as possible to the center-symmetric
\beqn
L \langle \Phi \rangle  = {\rm diag} \left( -\frac{2\pi [N/2]}{N},\,\,  -\frac{2\pi ([N/2]-1)}{N}, ....,\,
\frac{2\pi [N/2]}{N} \right) \,,
 \label{Eq:pattern2}
\eeqn
point, i.e.,  close to the vacuum of the pure  YM* theory in the weak coupling regime. 

At small $r(S_1)$, when  the gauge coupling is small, the eigenvalue fluctuations
around the center stabilized minimum (\ref{Eq:pattern2}) are small. This implies 
that at large distances
the gauge structure reduces to the ${\rm U}(1)^{N-1}$ subgroup of SU$(N)$. Due to 
the broken gauge symmetry, the  perturbative spectrum at low energies 
reduces to massless photons and massless fermions charged under the  ${\rm U}(1)^{N-1}$ gauge group.  We want to know whether or not
non-perturbative effects destabilize the masslessness of these excitations. More specifically, we want to understand whether or not the gauge fluctuations are gapped.  
As usual, non-perturbative  topological excitations are classified in powers of $e^{-S_0}$.

Appropriate analysis runs parallel to 
that in the chiral quiver theories.  It is slightly more technical, however.  
The differences  we will encounter are analogous to those
between the vector-like QCD(BF)* theory and QCD(AS)* theories discussed 
in our earlier work \cite{Shifman:2008ja}.   
We recall that  QCD(BF)*  was technically much simpler due to various implications of 
the Callias index theorem. In particular, in both QCD(BF)* and QCD(AS)*, 
there are $N$ monopole operators, but the former has a total of $2N$ 
zero modes distributed evenly (two for each monopole) 
between the monopoles. The latter has $2N-4$ 
zero modes, two for $N-2$ monopoles and nothing  for the remaining two.   
(See Eqs. (75) and (52) in 
\cite{Shifman:2008ja} and discussion on page 37 on the relation between the 
Callias and Atiyah--Singer index theorems.)  Also it is worth
recalling that for QCD(F)* theory with one fundamental fermion, 
the BPST instanton has two fermion zero modes.  
They must be distributed among $N$ monopoles as follows: 
two fermion insertions attached to one 
monopole, with no fermion insertions in the remaining $N-1$ monopoles. 
All these distributions of zero modes are a natural consequence of 
the Callias index theorem. 

Similar to what happens in the chiral quiver theories, 
in the ``traditional"
chiral theories $(N-1)$ out of $N$ monopole operators vanish due to averaging 
over a certain global part of the gauge symmetry. A single monopole operator 
(which does not carry  fermion zero mode insertions) is the only contribution in the   
$e^{-S_0} $ expansion at the level $e^{-S_0} $. 

Naive monopole operators 
can be found by truncating various monopole operators in the 
QCD(AS)*   and QCD(F)* theories of Ref. \cite{Shifman:2008ja}.  In a sense, 
the chiral theory at hand is a combination of QCD(AS)* and $(N-4)$ QCD(F)* 
stripped off of their chiral partners. The resulting operators are 
    \begin{eqnarray}
&& {\mathcal M}_{1}=   e^{ -S_{0}}  e^{i    \mbox{\boldmath $\alpha$} _1   \mbox{\boldmath 
$\sigma$} } \;  
( \lambda_1   + \lambda_2 )   \,,\nonumber\\[3mm]
&& {\mathcal M}_{2}=   e^{ -S_{0}}  e^{i   \mbox{\boldmath $\alpha$}_2    \mbox{\boldmath 
$\sigma$}  } \;  ( \lambda_2   + \lambda_3  )   \,,\nonumber\\[3mm]
&&\ldots \,,\nonumber\\[3mm]
&&{\mathcal M}_{m-1}=   e^{ -S_{0}}  e^{i  \mbox{\boldmath $\alpha_{m-1}$}    \mbox{\boldmath 
$\sigma$}   } \;  ( \lambda_{m-1}    + \lambda_{m}   ) 
\,,\nonumber\\[3mm]
&& {\mathcal M}_{m}=  e^{ -S_{0}}  e^{i  \mbox{\boldmath $\alpha$}_{m}    \mbox{\boldmath 
$\sigma$}   } \;
   ( 2 \lambda_{m}  )
\,,\nonumber\\[3mm]
&&  {\mathcal M}_{m+1}=  
 e^{ -S_{0}}  e^{i   \mbox{\boldmath $\alpha$}_{m+1}    \mbox{\boldmath 
$\sigma$}   } \;
  ( \lambda_{m}   + \lambda_{m-1}  )  
\,,\nonumber\\[3mm]
 &&\ldots  \,,\nonumber\\[3mm]
 && {\mathcal M}_{2m-2}=   e^{ -S_{0}}  e^{i   \mbox{\boldmath $\alpha$}_{2m-2}    \mbox{\boldmath 
$\sigma$}  } \;  
 ( \lambda_3   + \lambda_2  )    
\,,\nonumber\\[3mm]
&& {\mathcal M}_{2m-1}=   e^{ -S_{0}}  e^{i  \mbox{\boldmath $\alpha$}_{2m-1}    \mbox{\boldmath 
$\sigma$}   } \;  
   ( \lambda_2   + \lambda_1 )   
\,,\nonumber\\[3mm]
   && {\mathcal M}_{2m}=   e^{ -S_{0}}  e^{i   \mbox{\boldmath $\alpha$}_{2m}    \mbox{\boldmath 
$\sigma$}   }  \; (\psi_{I_1} \psi_{I_2} \ldots  \psi_{I_{N-4}} \epsilon^{I_1\ldots I_{N-4}}) 
 \; \,,
   \nonumber\\[3mm] 
    && {\mathcal M}_{2m +1}=   e^{ -S_{0}}  e^{i   \mbox{\boldmath $\alpha$}_{2m+1}    \mbox{\boldmath 
$\sigma$}   } \,.
\label{80}
           \end{eqnarray}
Obviously, none of these operators, 
except ${\mathcal M}_{2m +1}$, is invariant under the global 
gauge rotations. Hence, they all vanish. They are not even 
covariant with regards to the global gauge rotations. 
However, they are useful as building blocks, 
in building manifestly 
gauge and global symmetry invariant flux operators of the chiral theory. 

Let us pause here to make a remark 
regarding the ${\rm U}(1) \times {\rm SU}(N-4)$ invariance.  The latter is manifest, 
while the former is more tricky, 
and provides a consistency check. The invariance of the 
monopole operators under the global symmetry (\ref{U(1)}) requires the following 
continuous shifts for the $(N-1)$ varieties of the photons:
\begin{eqnarray}
&&  \mbox{\boldmath $\alpha$} _i   \mbox{\boldmath 
$\sigma$} \rightarrow  \mbox{\boldmath $\alpha$} _i  \mbox{\boldmath 
$\sigma$}  - (N-4) \alpha, \qquad i=1, \ldots N-2\,,
\nonumber\\[3mm]
&&  \mbox{\boldmath $\alpha$} _{N-1}   \mbox{\boldmath 
$\sigma$} \rightarrow  \mbox{\boldmath $\alpha$} _{N-1} \mbox{\boldmath 
$\sigma$}  +  (N-4) (N-2) \alpha \,.
\label{conttr}
\end{eqnarray}
If the chiral gauge theory will acquire a mass gap, 
this invariance must be consistent with the existence of the operator  
${\mathcal M}_{2m +1}$. As expected, 
\begin{eqnarray}
&&  \mbox{\boldmath $\alpha$} _{N}  
  \mbox{\boldmath 
$\sigma$} = - \sum_{i=1}^{N-1} 
 \mbox{\boldmath $\alpha$} _i  \mbox{\boldmath 
$\sigma$}  
 \rightarrow   \mbox{\boldmath $\alpha$} _{N}  
  \mbox{\boldmath 
$\sigma$}
\end{eqnarray}
where the shift of the photon  $\mbox{\boldmath $\alpha$} _{N-1}  \mbox{\boldmath  $\sigma$}$
cancels precisely the shifts of $N-2$ photons 
 $\mbox{\boldmath $\alpha$} _{i}  \mbox{\boldmath  $\sigma$}$ given in (\ref{conttr}).

In QCD(AS)*, there are both magnetic monopole operators and magnetic bion operators.  
As we discussed in the context of the chiral quiver theories, 
averaging over the global part of the gauge symmetry causes  
$(N-1)$ varieties of the monopole operators to vanish. Only 
$e^{ -S_{0}}  e^{i   \mbox{\boldmath $\alpha$}_{2m+1}    \mbox{\boldmath 
$\sigma$}   } $ does not vanish at this order. 
However, there are $N-1$ varieties of the photons,
and the operator  
${\mathcal M}_{2m +1}$   
renders only one linear combination massive. The major contribution 
to the mass term for the dual photons are due to magnetic bions
-- magnetically charged, but topologically null configurations which 
carry no fermionic zero modes.   In almost all 
chiral gauge theories, 
magnetic bions of various charges are abundant and are the main cause for the mass 
gap generation.  
These are all  $e^{ -2S_{0}}$ order effects.  The magnetic bions in our theory are  
 \begin{eqnarray}
&&{\mathcal B}^1_i : \Big( \frac{4\pi}{g}(\mbox{\boldmath $\alpha$}_i\, - \mbox{\boldmath $\alpha$}_{i-1}) 
\,, \; 0 \Big)  :  \qquad \;\; \;\; c_1 e^{-2S_0} e^{i ( \mbox{\boldmath $\alpha$}_i \, - \mbox{\boldmath $\alpha$}_{i-1} \,) \mbox{\boldmath $\sigma$}} \,,  \nonumber\\[3mm]
&& {\mathcal B}^{12}_{i,i} :  \Big(  \frac{4\pi}{g}(\mbox{\boldmath $\alpha$}_i\,   - \mbox{\boldmath $\alpha$}_{2m-i})\,, 0 \Big)  : \qquad \;\;
  c_2 e^{-2S_0} e^{i ( \mbox{\boldmath $\alpha$}_i -  
  \mbox{\boldmath $\alpha$}_{2m-i} ) \mbox{\boldmath $\sigma$} 
  } \, , 
  \nonumber\\[3mm]
  && {\mathcal B}^{12}_{i,i-1} :  \Big( \frac{4\pi}{g}( \mbox{\boldmath $\alpha$}_i\,   - \mbox{\boldmath $\alpha$}_{2m-i+1})\,, 0 \Big)  : \quad 
  c_2 e^{-2S_0} e^{i ( \mbox{\boldmath $\alpha$}_i -  
  \mbox{\boldmath $\alpha$}_{2m-i+1} ) \mbox{\boldmath $\sigma$} 
  } \, , 
  \nonumber\\[3mm]
&& {\mathcal B}^{12}_{i, i+1} :  \Big(  \frac{4\pi}{g}(\mbox{\boldmath $\alpha$}_i\,   
- \mbox{\boldmath $\alpha$}_{2m-i-1})\,, 0 \Big)  : \quad 
  c_2 e^{-2S_0} e^{i ( \mbox{\boldmath $\alpha$}_i -  
  \mbox{\boldmath $\alpha$}_{2m-i-1} ) \mbox{\boldmath $\sigma$} 
  } \, .
\end{eqnarray}
In the first line summation runs over $ i=1,\,\ldots,  \, 2m-1$
while in the second, third and fourth lines over $i= 1,\, \ldots,\, m-1$.  
The combined effect of the magnetic bions (which is of 
the order $e^{-2S_0}$) is
 \begin{equation}
 V_{\rm bion} (  \mbox{\boldmath $\sigma$} )   = m_W^3 g^{-6}  
 \left[ \sum_{i=1}^{2m-1} {\mathcal B}^1_i  + 
 \sum_{i=1}^{m-1} ( {\mathcal B}^{12}_{i,i} +  
 {\mathcal B}^{12}_{i,i+1} +  {\mathcal B}^{12}_{i,i-1} ) \right] + {\rm H.c.},
  \end{equation}
and the  monopole-instanton operator  
${\mathcal M}_{2m+1} $
(see Eq.~(\ref{80})) gives rise to the bosonic potential which renders all 
$N-1$ dual photons massive in the chiral SU$(N)$ theory implying Polyakov's confinement
in turn.  

The first global symmetry singlet operator 
which has multiple fermion zero mode insertions is also quite interesting. It appears  at the order  $e^{ -(N-1)S_{0}}$ in the  $e^{ -S_{0}}$ expansion.  In a sense,
it is a gauge singlet composite of the monopoles ${\mathcal M}_{1}, \ldots, {\mathcal M}_{2m} $.  In other words,  it is 
an object whose action is $e^{ -(N-1)S_{0}}$ and whose quantum numbers 
are the same as those of   
``instanton minus the monopole"   ${\mathcal M}_{2m+1} $,
\begin{equation}
I_{\rm  instanton}\,\, \overline {\mathcal M}_{2m+1}  \,.
\end{equation}
A variant of this topological object was previously 
identified in a vector-like context in \cite{Brower:1996js}.
Naturally, the instanton of the four-dimensional theory (which shows
up  in 
the order  $e^{ -NS_{0}} $) 
can be thought of as a composite of $N$ types of monopoles. 

This describes dynamics of the chiral SU$(N)$ theory 
with one AS and $N-4$ fundamental Weyl fermions at small $r(S_1)$.  
Note that the ``conventional" chiral theories we have just discussed
have a number
of continuous non-anomalous chiral symmetries. 
For instance, in the simplest example, 
SU(5) with one decuplet $X^{[ij]}$ and one antiquintet
$V_i$, we have a continuous chiral U(1) generated
by the current $\bar{X}_{\dot\alpha} X_\alpha
- 3 \bar{V}_{\dot\alpha} V_\alpha$.
As we saw, at small $r(S_1)$ this chiral U(1) symmetry remains unbroken.
Thus, we have an example of a Yang--Mills theory {\em with} confinement,
but {\em no} chiral symmetry breaking. This is also valid for ${\rm SU}(N)$ gauge theories with 
$N \geq 6$. 

What happens as we pass to large $r(S_1)$ (eventually, $r(S_1)\to\infty$)?  
In $R_4$, there are two complementary description of these class of theories
\cite{Dimopoulos:1980hn,Peskin:1982mu}:

\vspace{1mm}

{\bf The Higgs picture:} The gauge  and chiral symmetry realization are
\begin{equation}
{\rm SU}(N) \times [{\rm U}(1) \times {\rm SU}(N-4)]_{\rm flavor} \longrightarrow 
{\rm SU}(4)  \times  [{\rm U}(1) \times {\rm SU}(N-4)]'_{\rm flavor} 
\end{equation}
where the $ [{\rm U}(1) \times {\rm SU}(N-4)]'_{\rm flavor}$ symmetry is realized in  the diagonal group of color and flavor.  Non-vanishing color-flavor locked (non-singlet)
chiral condensates appear.

{\bf The confinement (symmetric) picture:} None of the symmetries are broken,
but there exist massless composite {\em fermions} which are bound by confining ${\rm SU}(N)$ forces and which satisfy non-trivial 't Hooft matching conditions.  

\vspace{1mm}

The massless spectrum, the global anomalies -- the anomaly generated 
by three ${\rm SU}(N-4)$ currents, 
by three  ${\rm U}(1)$ currents,  and the mixed anomaly 
generated by one  ${\rm U}(1)$ and 
two ${\rm SU}(N-4)$ currents -- and the unbroken global symmetry of these two descriptions match, although the underlying descriptions of dynamics are quite different, as explained in 
\cite{Dimopoulos:1980hn,Peskin:1982mu}.
   
As we pass to large $r(S_1)$,  in the sense of conventional order parameters, and according to the Landau--Ginzburg--Wilson paradigm of phase transitions,  there seems to be no sharp distinction   
{\em en route} between physics of small-$r(S_1)$ and large-$r(S_1)$ theory.  
Moreover, our description  of the small-$r(S_1)$  physics seems to be a natural continuation of 
the confinement picture.  However, we suspect that the non-perturbative spectrum of these theories 
will have some unusual aspects on the way.    We plan to address this issue in a separate publication.

\section{Volume independence of  chiral  theories  in the $N=\infty$ limit}
\label{victni}

The  large-$N$ limit of YM theory formulated 
on $R_{4-d} \times T_d$  is independent of the volume of the $d$-torus $T_d$, provided 
the center symmetry is not 
broken~\cite{Eguchi:1982nm,Yaffe:1981vf, Bhanot:1982sh}. We will refer to this 
non-perturbative property 
of the gauge theories as volume independence. 
If we take just one dimension compactified, volume independence  translates into 
temperature independence   in the center symmetric, confined phase. The 
Eguchi--Kawai (EK) reduction \cite{Eguchi:1982nm},  relating an infinite volume lattice gauge theory to a single-site matrix model is another  special case of large-$N$ volume independence. 
   
Unfortunately, for {\it all} asymptotically free confining gauge theories formulated on 
$R_3 \times S_1$, with $S_1$ being a thermal circle, the center symmetry does break spontaneously below a critical size $L_c$ (the deconfined phase) invalidating the EK reduction.  A way to preserve the volume independence in arbitrarily small volumes in 
confining YM and QCD-like theories (with vector-like fermions)
is through deforming  \cite{Unsal:2008ch} YM or QCD (passing to YM* or QCD*,
respectively)  by adding the center stabilizing deformation potential  (\ref{fourma}).
In the $N \rightarrow \infty$  limit, pure YM or QCD 
theories with one- or two-index representations on $R_4$ are equivalent to deformed  
YM* and QCD* theories on $S_1 \times  R_3$ regardless of the size of $S_1$.
Since YM* and QCD* theories satisfy the full volume independence, they 
 provide  a reduced model of SU$(\infty)$ YM and QCD  on 
 $R_4$.\footnote{For a long time, it was thought  that there are only two working schemes to preserve the volume  independence in the $N=\infty$ theories, called quenching \cite{Bhanot:1982sh} and twisting \cite{GonzalezArroyo:1982ub}. Unfortunately, both schemes were  recently shown to fail at weak coupling.  The quenching scheme is insufficient to stabilize the center symmetry breaking down to a diagonal subgroup \cite{Bringoltz:2008av}, i.e.,   the Wilson lines 
 in different directions are locked. 
The twisting scheme 
fails due to entropic effects, as shown in Ref.~\cite{Azeyanagi:2007su,Teper:2006sp}.   
Very recently, it was shown that QCD with multiple adjoint fermions with {\it periodic} spin connection (i.e., a non-thermal, zero temperature compactification with solely quantum fluctuations as opposed to thermal) 
satisfies the full volume independence without any modification of the measure (as 
is done in quenching) or the action (as is done in twisting) \cite{Kovtun:2007py}.  
The physical difference between the thermal and non-thermal compactification 
arises due to  sharp  qualitative distinction between  the   thermal and quantum 
fluctuations in gauge theories. 
The quantum fluctuations induced by adjoint  fermions are sufficient 
to dynamically stabilize the center-symmetric vacua. 
The double-trace deformation is inspired  by this fermion-induced stabilization.} 

\begin{figure}
\begin{center}
\includegraphics[angle=-90,width=4 in]{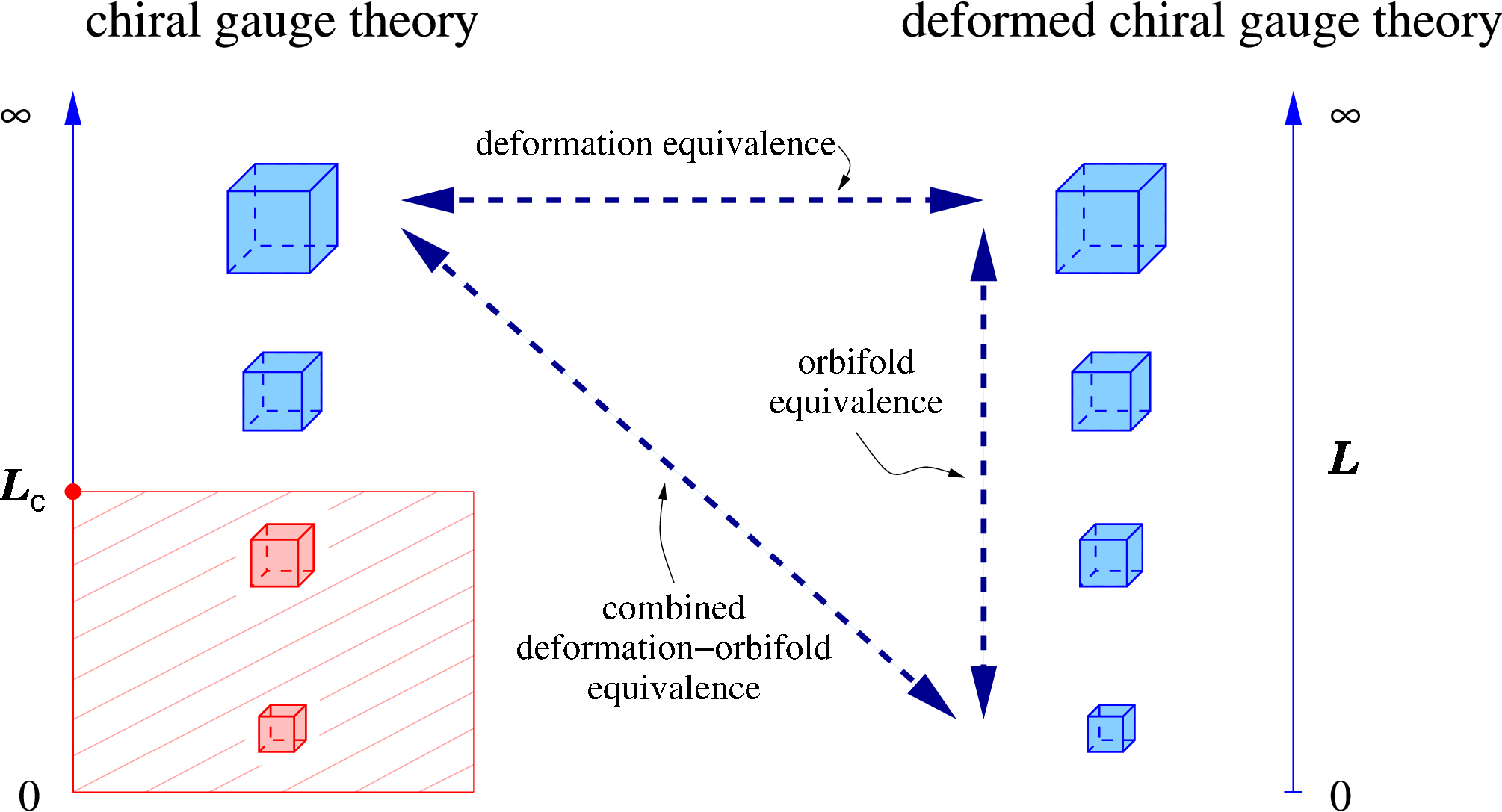}
\caption
    {\small The SU$(\infty)$  deformed chiral theory, unlike the original chiral theory which is 
    expected to possess a center symmetry changing transition at $L_{\rm c} \sim \Lambda^{-1}$,  
    satisfies full volume independence down to arbitrarily small volumes.  
  The figure is adapted from 
 Ref.~\cite{Unsal:2008ch}. }
 \label{reduced1}
\end{center}
\end{figure}

Here, we will propose  a generalization of  the large-$N$ volume independence for 
strongly coupled chiral gauge theories. The idea is simple and, in essence, the same as 
that in QCD-like theories with vector-like matter \cite{Unsal:2008ch}. The main 
idea of our suggestion is   depicted in 
Fig.~\ref{reduced1}.
We do not know yet  whether or not this may have practical (numerical) or analytical utility. 
Given that the lattice implementation of non-Abelian chiral gauge 
theories is still far from being settled \cite{Golterman:2004qv}, 
we can only hope that our construction could be useful for numerical purposes in the future. 

The chiral $[{\rm SU}(N)]^K$ gauge theories formulated on $S_1 \times R_3$ possess 
a $[Z_N]_C$ center symmetry, and the cyclic $Z_K$ symmetry of the quiver
(see Sect.~\ref{cogtg}).  
As was discussed in Sect.~\ref{srrwtdtd}, the  $[Z_N]_C$ symmetry 
spontaneously breaks at small $r(S_1)$. 

We add $[Z_N]_C$ stabilizing  $Z_K$ singlet deformation potential given in (\ref{fourma}).  
In the $N=\infty$ limit,  the chiral theory on $R_4$ is equivalent to the deformed chiral quiver theory on  $S_1 \times R_3$ for any $r(S_1)$.  

If  we compactify  $R_4$ down to  
$R_{4-d} \times T_d$ and stabilize the center symmetry $([Z_N]_C)^d$ 
in the small-$T_d$ regime by a deformation potential,  
\begin{equation}
    P[U_{J,1}, \cdots, U_{J,d} ] \equiv 
    A \frac{1}{\pi^2 L^4}
 \sum_{J=1}^{K}   \sum_{(n_1, \cdots, n_d) \in (\Z^d - {\bf 0}) }^{\infty}
    \frac
    {\left|\tr\left(U_{J,1}^{n_1} \cdots U_{J, d}^{n_d} \right)\right|^2}
    {\left(n_1^2 + \cdots + n_d^2\right)^2}
    \,,
\label{eq:potential2}
\end{equation}
physics of the chiral theory satisfies volume independence.  
The proof of volume independence is  along the same lines 
as in Ref.~\cite{Unsal:2008ch}.   

If a confining asymptotically free gauge theory preserves its center symmetry on 
$S_1 \times R_3$, or, in 
general, on $R_{4-d} \times T_d$, then there are two ways  
to take the decompactification limit. The first is to take $r(S_1) \rightarrow \infty$ 
while the second is to take $N\rightarrow \infty$ at a finite  $r(S_1)$. The latter is a manifestation of the large-$N$ volume independence, or (a working) EK reduction.   
 
 \subsection{Neutral sector (untwisted) observables in
 the  EK\\
 reduction} 
 
 The volume independence can be formulated, as 
 shown in \cite{Kovtun:2007py,Unsal:2008ch}, as an orbifold equivalence of theories related to one another by orbifold 
 projections.  The operators which are invariant under the symmetries used in projections 
 constitute a neutral sector (called untwisted sector in string theory). The dynamics of the parent and daughter theories in  their  neutral sector coincide in the large-$N$ limit, provided
 the symmetries defining the neutral sectors are not spontaneously broken.

 For volume independence to be valid in chiral gauge theories, it is necessary (and sufficient) that the center symmetry and lattice translation symmetries remain unbroken (spontaneously). 
 One  point that we wish to emphasize is  
 that the large $N$ equivalence only applies to a sub-sector  
 of a gauge theory  and not the whole theory. Below, we will give few simple examples of 
observables
 that can be extracted in certain limiting cases of the volume independence. 
  
Perhaps, the most famous example of the volume independence  is the Eguchi--Kawai  reduction  
which goes all the way to a matrix model.  As stated above, this works in the strong 
coupling phase of the lattice gauge theory, 
but tempered by a phase transition in the phase continuously connected to continuum limit.   
Our deformation prevents such breaking in the continuum; 
we can have a full reduction in terms of our deformed theory. 
In this  example,   
the large-$N$ reduction relates physical quantities in the zero momentum sector of the  
original  theory to the observables in the reduced deformed theory. 
 This means, we can extract the expectation 
values of the operators, such as Wilson loops of arbitrary size and shape, 
 \begin{equation}
\Big\langle W[C]\Big\rangle_{R_4} = \Big\langle W[C] \Big\rangle_{\rm deformed \;  reduced} 
  + O(1/N^2) \,,
  \label{wilson}
\end{equation}
and thermodynamic 
quantities such as free energies, pressure or heat capacities.  In Eq.~(\ref{wilson}), if the 
reduced theory is defined on an  arbitrarily small $T_4$ (or a point in a lattice formulation), 
then the image of the original Wilson loop is a 
multi-winding string, neutral under the center symmetry.   
Although defined on a point, the expectation value of such a 
Wilson loop will obey the area 
law of confinement, with an identical string tension as in the 
theory on $R_4$.  At a formal level (i.e. in
the Schwinger--Dyson equations for the Wilson loops) 
this can be seen 
through the fact that both the theory on $R_4$ and the deformed reduced theory on $T_4$ satisfy identical set of loop equations as long as the center symmetry is unbroken in the 
latter \cite{Eguchi:1982nm}.  Our  deformed chiral gauge theories are   
center-symmetric, by construction.   

In order to access  the non-perturbative 
spectrum of the gauge theory, it is necessary to calculate long-distance connected correlators. 
Hence, the excitation spectra is not part of the neutral sector observables in the fully reduced theory.  To render the spectra a part of the neutral sector accessible to the reduced model, 
one needs to keep  at least one dimension non-compact. 
The resulting reduced  model is an SU$(\infty)$ matrix quantum mechanics, a theory defined on 
$R \times \{\rm point\}$.

For example, let $(x_0, {\bf x}) \in R \times R_3$.  Then, the large-$N$ 
volume independence of the  deformed chiral  theory and the 
combination of the large-$N$ deformation-orbifold equivalences indicated by the diagonal 
arrow in  Fig.~\ref{reduced1}  implies 
\begin{equation}
\Big\langle \tr F^2( x_0, {\bf x})  \, \tr F^2(x_0', {\bf x} ) \Big\rangle_{\rm con., \; R_4} = 
\Big\langle \tr F^2(x_0)  \, \tr F^2(x_0') \Big\rangle_{\rm con., \; R} 
  + O(1/N^2) 
\end{equation}
at any separation   $|x_0 - x_0'|$. In particular, assume $|x_0 - x_0'|\gg \Lambda^{-1}$. Then,
we can extract the non-perturbative spectrum of the SU$(\infty)$  chiral gauge theory by studying the non-perturbative spectrum of the deformed  
 matrix quantum mechanics. Let ${\cal H}$ denote the Hilbert space of the corresponding 
 gauge theory. Then 
 \begin{equation}
{\rm Spec} [{\cal H}]^{ \rm chiral, R_4} = {\rm Spec} [{\cal H}]^{\rm deformed \; \; reduced, R } \,, \,\,\,\,{\rm at} \,\,\,\,  N=\infty\,.
\end{equation}
The equality of the  non-perturbative spectra is a consequence of 
the large-$N$ volume independence which demands
\begin{equation}
\frac{\partial}{\partial L} {\rm Spec} [{\cal H}]^{ {\rm deformed}, R_{4-d} \times T_d}(L) = O\left(\frac{1}{N^2} \right) \longrightarrow 0
\end{equation}
in the non-Abelian confinement regime.  
  It is certainly desirable to study the reduced matrix quantum mechanics in detail, and see whether or not it is  more tractable than the gauge theories on $R_4$.  We currently 
  have no idea whether the SU$(\infty)$ reduced model is easier than the chiral theory on $R_4$.
However, the importance of the new formulation is self-evident, especially   considering that 
 lattice construction of chiral theories is still impractical.

 \subsection{Refined (Abelian) large-$N$ limit}
\label{ralnl}
 
Confinement on $R_4$ in the chiral gauge theories 
considered is believed to be non-Abelian. By this we mean that
there is no length scale at which the large-distance theory can be described by 
dynamics in the maximal Abelian subgroup. The volume independence 
implies that for finite $r(S_1)$  dynamical 
Abelianization  $ [{\rm SU}(N)]^K  \longrightarrow [{\rm U}(1)^{N-1}]^K$ 
does not occur in the $N=\infty$ limit. At this point we need to explain 
how our fixed-$N$, small-$r(S_1)$ analysis fits together 
with the volume independence in the large-$N$ limit. In other words, we need to 
elucidate the issue of the domain of validity of our general analysis 
in which large-distance dynamics of the chiral gauge theory 
can be analytically described by the maximal Abelian subgroup  
$[{\rm U}(1)^{N-1}]^K$.

\begin{figure}
\begin{center}
\includegraphics[angle=-90,width=3 in]{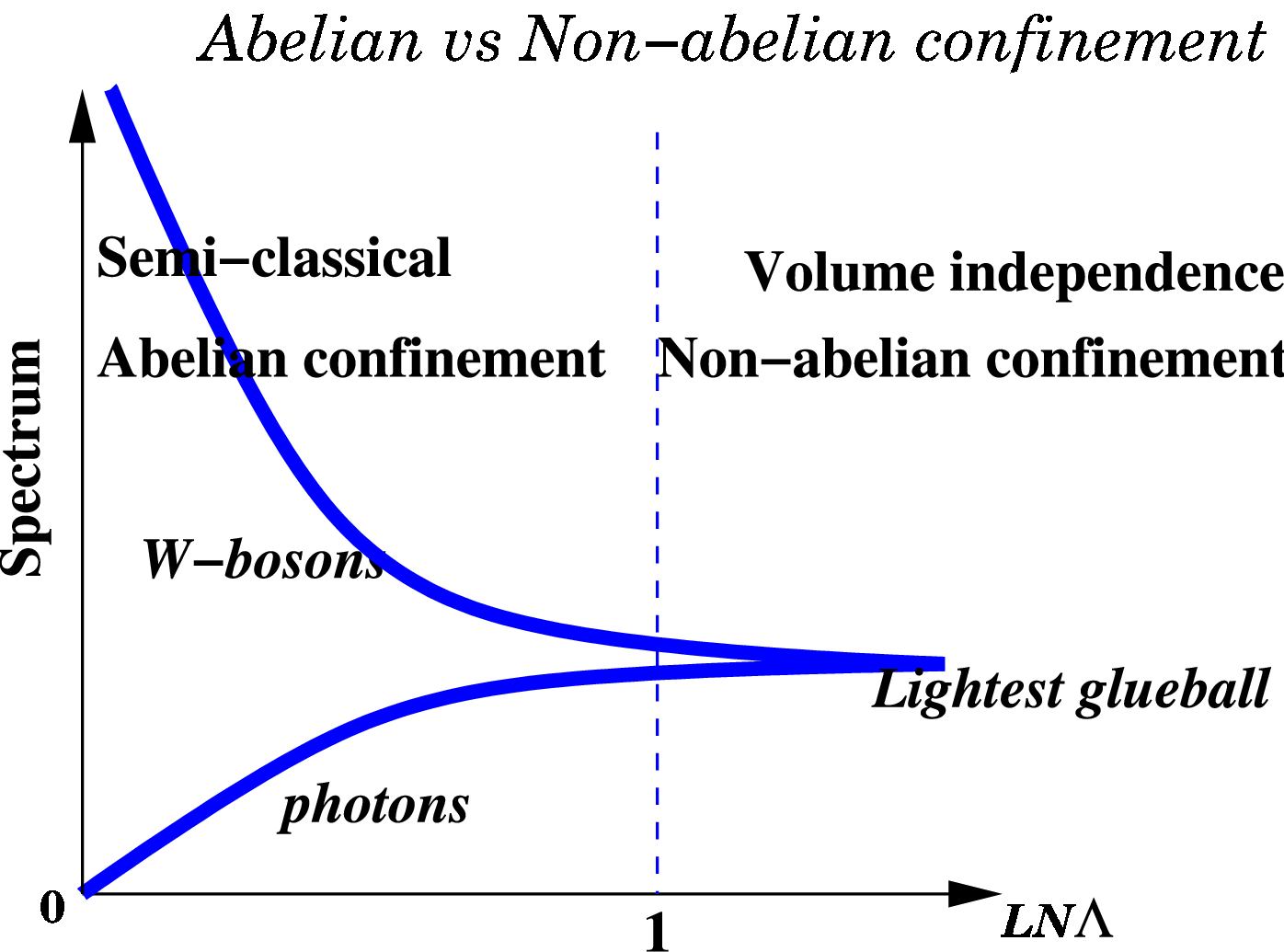}
\caption
    {\small  Cartoon of the spectral properties of the lightest glueball
     as a function of $LN \Lambda$.  At small 
    $LN \Lambda$,     
         a semi-classical   analysis is possible and reliable. In this regime,
        Abelian confinement is operative. 
 As  $N\rightarrow \infty$, this window shrinks to zero  in  $r(S_1)$ and volume independence 
 (and non-Abelian confinement) takes over  at any finite $r(S_1)$. 
   }
 \label{saturation}
\end{center}
\end{figure}

The quasiclassical analysis of the deformed chiral theories is reliable as long as there is a 
parametric separation of scales between photons (which are perturbatively
massless) and the lightest $W$ bosons of the spontaneously broken non-Abelian theory. 
The non-perturbative photon mass and that of the lightest $W$ bosons are 
  \begin{eqnarray}
  m_{\sigma} \sim e^{-S_0}  \sim \Lambda( \Lambda  N L)^2, \qquad m_W = \frac{2 \pi}{LN} \,,
  \label{tgv1}
  \end{eqnarray}
where we expressed the photon mass in the units of $\Lambda$.  
The photon mass is an increasing function of $LN$ while the lightest $W$-boson mass is a decreasing function as shown in Fig.~\ref{saturation}.  As long as the ratio of the two masses is smaller than unity, the large-distance dynamics can be accurately described by photons in the maximal Abelian subgroup. This implies 
    \begin{eqnarray}
 \frac{ m_{\sigma}}{m_W} \sim ( \Lambda  N L)^3 \ll 1\,.
 \label{pmas}
  \end{eqnarray}
At $LN\Lambda \sim 1$ the separation of scales is lost; we can no longer describe 
large-distance physics limiting ourselves to photons and light fermions.
The theory passes from Abelian to non-Abelian confinement. 
At $LN\Lambda \gg 1$ the theory lacks a weak coupling description regardless of how small $r(S_1)$ is, despite asymptotic freedom and despite the fact that we imprisoned the gauge theory in an arbitrarily small box. In a sense, the effective infrared cut-off
in the case at hand is $LN$ rather than $L$. Of course,  this statement is the essence of the concept of volume independence. 

Thus, we see that everything fits together very well. 
As we make $N$ larger, the domain of validity of our analysis shrinks as $L \ll {1}/(\Lambda  N)$. This is how 
the volume independence and quasiclassical analysis are intertwined, 
so that both hold without invalidating each other.  
Note that, were it not for the nontrivial $N$ dependence in Eq.~(\ref{pmas}),
our analysis would come in contradiction with
the large-$N$ volume independence.  Hence, the scale  $
r\sim {1}/(N\Lambda)$ is an important physical scale in gauge theories --
this is the scale where Abelian confinement gives place to 
non-Abelian confinement. 
  
The above discussion implies that a {\em refined (Abelian) large-$N$ limit} 
might exist in which the combination 
  \begin{equation}
  L N \Lambda = \epsilon \ll1  
  \end{equation}
is kept fixed and small (we must keep $\Lambda$ fixed too, of course).     
If the large-$N$ limit is taken according to this double scaling, physics can be 
described by a (compact) Abelian 
 $ [{\rm U}(1)^{\infty}]^K $ refined  gauge  structure
at large distances.

In the 
$N=\infty $ limit, the implication of the volume independence is much stronger.  
It implies that all non-perturbative 
features, such as the glueball  spectrum, string tensions, chiral condensates, etc.,
are independent of $r(S_1)$. At finite $N$, physics at $L \leq {1}/(N\Lambda)$  
has an $L$ dependence. The fact that the chiral condensate came out to be 
independent of $L$ even at small $N$ was due to the  relation between 
$e^{-S_0}$ and the strong scale at the one-loop level. 
This might seem as a welcome one-loop beta-function accident. It is not, given that
the $Z_K$ orbifold theories are perturbatively planar equivalent to
SYM theory \cite{Kachru:1998ys,stras1}, and the perturbative equivalence 
implies coinciding renormalization group $\beta$ function and the same strong scale by dimensional transmutation.  
We believe that the chiral condensates  in the $Z_K$ orbifold theories are saturated by appropriate flux-ring operators at small $r(S_1)$, just like the gluino condensate is 
saturated by the monopole operators in $\N=1$ SYM theory \cite{Davies:1999uw}, and 
QCD(BF/AS/S)* \cite{Shifman:2008ja}. \\

\mbox{}
\\
\mbox{}
  
\section{Conclusions}

The double-trace deformation gives us a controllable dynamical framework to study 
non-perturbative aspects of the strongly coupled chiral 
gauge theories. Our analysis is valid at small $r( S_1)$. Due to the absence of 
a confinement-deconfinement phase transition in the deformed theories our analysis
must be qualitatively valid in the chiral theories on $R_4$.  We established, 
at small $r(S_1)$, the existence of the dual photon masses 
generated by bions, and, hence,  linear confinement. We calculated chiral
condensates which determine the pattern of 
discrete $\chi$SB and the number of distinct vacua.
The form of these condensates is in agreement with what one would
naively guess assuming ``minimality" and gauge invariance. 
At small $r(S_1)$ they are generated by ring operators.

At small $r(S_1)$, reduction of the full gauge symmetry down to 
the  maximal Abelian subgroup (e.g.  $[{\rm U}(1)^{N-1}]^K$
in the quivers) occurs in the deformed chiral 
theories.  One of surprising findings of our work
 is the 
vanishing of the monopole operators.   In other words, despite the gauge symmetry breaking, 
{\em the monopoles  per se do not contribute to
non-perturbative dynamics.}  The leading non-perturbative effects are due to 
various flux operators, such as magnetic bions and magnetic ring operators
which may be thought of as molecules  built of the  monopole-instantons. 
This is in a striking difference with YM* and QCD* theories with vector-like matter where the monopole effects are crucial.  
 
Our work also shows that  non-perturbative effects do not lead to
inconsistencies in 
dynamics of the  chiral gauge theories.  
In this sense, our suggestion provides an additional argument in favor of non-perturbative 
consistency of the chiral gauge theories.  

We outlined a reduced matrix model for the large-$N$ chiral gauge theories, along the lines of 
working EK reductions. The main lesson here is that non-perturbative aspects (such as the spectrum)  
of the $N=\infty$ chiral theory on $R_4$ 
are identical to those of the reduced deformed theory on $R_{4-d} \times T_d$ where $T_d$ is a 
 $d$-dimensional torus. A special case is a very small $T_3$ implying
 that  non-perturbative spectrum of the chiral theory can be deduced by studying $N=\infty$ 
matrix quantum mechanics.  It would be instructive to study such 
quantum-mechanical systems. 
  
We also remarked that the large-$N$ volume independence and the existence of the volume 
dependent quasi-classical regime on $S_1 \times R_3$ 
with small $r(S_1)$
are not in contradiction with each other, 
due to non-trivial region of validity of the latter, i.e.,   $r(S_1) N \Lambda \ll 1$. In the 
small-$r(S_1)$ regime, 
Abelian confinement is operative. The  volume independence is a non-perturbative property of the non-Abelian confinement regime.  In our opinion, currently, the most important  question  in vector-like QCD* theories and deformed chiral gauge theories is to understand the transition from the Abelian to non-Abelian confinement regimes in the vicinity of 
$r(S_1)\sim  1/(N \Lambda)$. The importance of this regime is due to volume independence. 
The physical observables of gauge theories on $R_4$ do get saturated by non-perturbative 
dynamics above the $r(S_1)\sim  1/(N \Lambda)$ scale.  After non-perturbative 
saturation takes place, the observables between the finite and 
infinite $r(S_1)$  theories can only differ by small  $O(1/N^2)$ effects.   
 
\section*{Acknowledgments}
\addcontentsline{toc}{section}{Acknowledgments}

We thank E. Poppitz for sharing with us his 
unpublished notes on chiral determinants, and useful 
remarks on the paper.  M.S. is grateful to G. Korchemsky and A. Vainshtein
for discussions.
M.\"U. thanks S. Dimopoulos,  M. Peskin,  E. Poppitz, M. Golterman for illuminating 
conversations about  chiral gauge theories.  We thank the Galileo Galilei Institute for
Theoretical Physics in Florence for their hospitality and INFN for partial support at the final stages
of this work. The work of M.S. is supported in part by DOE grant DE-FG02-94ER40823 and
by {\em Chaire Internationalle de Recherche Blaise
Pascal} de l'Etat et de la R\'{e}goin d'Ille-de-France, g\'{e}r\'{e}e par la 
Fondation de l'Ecole Normale Sup\'{e}rieure.
The work of  M.\"U. is supported by the U.S.\ Department of Energy Grant DE-AC02-76SF00515.

\end{document}